\documentclass[aps,prx,twocolumn,showpacs,showkeys,footinbib,superscriptaddress]{revtex4-2}

\usepackage{amsmath}
\usepackage{amssymb}
\usepackage{graphicx}
\usepackage{bm}
\usepackage{color}
\usepackage{relsize}
\usepackage{braket}
\usepackage[caption=false]{subfig}
\usepackage{dcolumn}

\usepackage{hyperref}
\usepackage[all]{hypcap}

\usepackage{tabularx}

\begin{document}
\title{Landau-level composition of bound exciton states in magnetic field}
\date{\today}

\author{Dinh Van Tuan}
\email[]{vdinh@ur.rochester.edu}
\affiliation{Department of Electrical and Computer Engineering, University of Rochester, Rochester, New York 14627, USA}
\author{Hanan~Dery}
\affiliation{Department of Electrical and Computer Engineering, University of Rochester, Rochester, New York 14627, USA}
\affiliation{Department of Physics and Astronomy, University of Rochester, Rochester, New York 14627, USA}

\begin{abstract} 
We present a theory that studies the state composition of a bound exciton in magnetic field.  Using a basis set made of products of free electron and hole wavefunctions in Landau gauge, we derive a secular equation which shows the relation between Landau levels (LLs) of the electron and hole when a bound exciton is formed.  Focusing on excitons in the light cone, we establish a scattering selection rule for the interaction of an electron in LL $n_\text{e}$ with a hole in LL $n_\text{h}$. We solve the resulting secular equation and identify a simple pairing law, $n_\text{e} = n_\text{h} + l$, which informs us on the construction of a bound exciton state with magnetic quantum number $l$, and on the interaction of the exciton magnetic moment with magnetic field. We obtain good agreement between theory results and recent measurements of  the diamagnetic shifts of  exciton  states  in WSe$_2$ monolayers.
\end{abstract}


\maketitle

\section{Introduction}

Magneto-optical spectroscopy is a powerful tool to investigate physical properties of materials based on the applications of light excitation and magnetic field  \cite{Sato_FPhys22}.  The light excitation helps to reveal molecular structures, band-gap energies, and the electronic band structure, while the magnetic field helps to extract further information through its interaction with the spin, valley, and orbital angular momenta. Magneto-optical spectroscopy studies of  two-dimensional (2D) materials are especially interesting on account of the enhanced Coulomb interaction due to their reduced dimensionality. The enhanced interaction results in tightly bound electron-hole pairs (excitons), which manifest through salient excitonic resonances in the optical spectrum. Transition-metal dichalcogenide (TMD) monolayers have been a recent focus  of studies of excitons in magnetic field \cite{Stier_PRL2018, Liu_PRL20,Molas_PRL19,Goryca_NanoLett22,Li_PRL20,Wang_PRX20,Wang_NanoLett20,Smolenski_PRL2019,Goryca_NatCom19,Smolenski_PRL2022,Li_NatComm20,Spiridonova_PLA20,Liu_PRB19,Pico_PRB25,Have_PRB19}, wherein  resident carriers (conduction-band electrons or valence-band holes) are hosted in valleys with opposite spin configurations and interact differently with magnetic field \cite{Li_PRL14,Srivastava_NatPhys15,Wang_2DMat15,Aivazian_NatPhys15,MacNeill_PRL15,Mitioglu_NanoLett15,Arora_NanoLett16,Stier_NatComm16,Koperski_2DM19,Robert_PRL21,He_NatComm20,Ren_PRB23}.  The ability of the magnetic field to spin and valley polarize the resident carriers has become a useful tool to explore interactions between excitonic states and charge particles with different quantum numbers \cite{Li_PRL20,Liu_PRL20,Goryca_NanoLett22,VanTuan_PRL22,Choi_PRB23}.

To date, magneto-optical selection rules of materials with Dirac Hamiltonians were obtained for optical transitions between Landau levels (LLs) of a free electron-hole pair \cite{Rose_PRB13,Tabert_PRL13,Tabert_PRB13,Li_PRB24,Wang_PRB22}. However, whereas the motion of a free resident carrier in magnetic field is described by LL quantization, a  charge-neutral exciton made of a tightly bound electron-hole pair is inert to Lorentz force \cite{Whittaker_PRB97}.    This contrast raises the question: How can a bound exciton state in magnetic field be  expressed  through LL quantization of its electron and hole components? Although magneto-excitons have been investigated for more than half a century  \cite{KnoxBook, Sugawara_PRB93, Hou_PRB91,Reynolds_APL94,Bugajski_SSC86,Someya_PRL95,Bayer_PRB98,Walck_PRB98,MiuraBook,Lerner_JETP80,Lozovik_PRB02},    the answer to this question is not well-understood.  Recently, an effort along this direction has been attempted to explain the unusual spin properties of InP wurtzite nanowires \cite{Tedeschi_PRB19,Junior_PRB19}. However, the method is limited to several LLs for the electron and hole components, making it inapplicable for materials with large exciton binding energy like TMDs \cite{Tedeschi_PRB19}. Furthermore,  since the energy of an exciton varies when its magnetic moment  interacts with magnetic field \cite{Lamb1952,KnoxBook,MiuraBook,Cong_EMO18,Donck_PRB18,Raczynska_NJP19,Gorkov_JETP68,Whittaker_SSC88, Katsch_PRB20},  it also elicits the question: what type of relation between LLs of the electron and hole produces the exciton magnetic moment?

We address these questions by using a basis set made of products of free electron and hole wavefunctions in Landau gauge.  Using this basis set, we derive Coulomb matrix elements and study the spectrum of excitons in magnetic field through a Hamiltonian that solely depends on LL indices of its electron and hole states. For an exciton in the light cone, the resulting Hamiltonian has a built-in selection rule for the scattering between LLs of the electron and hole. Interestingly, the magnetic quantum number of the exciton is established via a pairing law between the electron and hole LLs, from which we devise an elegant way to understand the exciton magnetic moment and its interaction with the magnetic field.  Our  results on diamagnetic shifts of magneto-excitons in WSe$_2$ monolayer are in good agreement with experimental data.  


This paper is organized as follows. Section~\ref{sec:Mov} provides a brief review of the free electron (or hole) wavefunctions in Landau gauge. These wavefunctions are then used to derive the Coulomb matrix element and exciton Hamiltonian in Sec.~\ref{sec:theory}, where we introduce the scattering selection rule for the electron-hole interaction when the exciton resides in the light cone. The selection rule leads to a pairing law between LLs of the electron and hole components.  Section~\ref{sec:results} provides numerical results of the LL compositions of magneto-excitons in their ground and excited states. We compare our numerical results with previously calculated results that were derived from a real-space Hamiltonian of an exciton in magnetic field. We  discuss the exciton magnetic moment, compare our results on diamagnetic shifts with experimental data,  and analyze exciton states with finite center-of-mass (CoM) momentum.  A summary and outlook are given in Sec.~\ref{sec:con}. The appendices include detailed derivations, validation of the model, and parameters used in numerical  calculations.

\section{Background}\label{sec:Mov}
We start by considering a free electron with electric charge $-e$ (or free hole with $+e$) in a 2D sample whose area is $L_x \times L_y$. The sample is subjected to an out-of-plane magnetic field $\bf B\,$$=\,B{\hat{\bf z}}$.   Using Landau gauge, the vector potential is chosen as \cite{Landau_30,VignaleBook}
\begin{equation}
{\bf A} =- x B {\hat{\bf y}},
\label{Eq:LandauGauge}
\end{equation}
and the resulting Hamiltonian of the free electron (hole) reads
\begin{equation}
H_0 = \frac{1}{2m} \left( p_x^2 + \left(p_y \mp eBx \right)^2 \right). \label{eq:H}
\end{equation}
The upper (lower) sign denotes the electron (hole) case, and $m$ is the effective mass. Since $[H_0,p_y] = 0$, the wavevector $k_y$ along the $y$ direction is a good quantum number (constant of motion), and the eigensolution of the Hamiltonian can  be written as
\begin{equation}
\psi_{k_y}(x,y) = \frac{e^{ik_yy}}{\sqrt{L_y}}f_{k_y}(x). \label{Eq:Eivec}
\end{equation}
The motion along the $x$ direction is  satisfied by the one-dimensional equation of the quantum harmonic oscillator
\begin{equation} 
\left( \frac{p_x^2}{2m} + \frac{m \omega^2}{2} \left( x \mp k_y \ell_\text{B}^2 \right)^2\right) f_{k_y}(x) =\varepsilon_{k_y} f_{k_y}(x),
\end{equation}
oscillating about the equilibrium position  
\begin{equation}
 x^* = \pm k_y \ell_\text{B}^2.
 \label{Eq:equiPos}
\end{equation}
$\ell_\text{B} = \sqrt{\hbar/(eB)}$ is the magnetic length and $\omega = eB / m$ is the cyclotron frequency.  The eigenenergies are quantized in LLs  
\begin{equation} 
\varepsilon_{n,k_y} \equiv \varepsilon_n= \hbar \omega \left( n+ \frac{1}{2} \right),
\label{Eq:EigEner}
\end{equation}
and their corresponding eigenstates are \cite{Landau_30, VignaleBook}
\begin{equation} 
\langle {\bf r}| n,k_y \rangle  \equiv \psi_{n,k_y}(x,y) =  \frac{e^{ik_yy}}{\sqrt{ L_y \ell_\text{B}  }}  \,\, \tilde{H}_n\left( \frac{ x \mp k_y \ell_\text{B}^2}{\ell_\text{B}} \right).
 \label{Eq:LandauWave}   
\end{equation} 
$\tilde{H}_n(x)$ are orthonormal Hermite-Gaussian functions, constructed from Hermite polynomials $H_n(x)$ and a Gaussian weight function (Appendix \ref{ap:p_wave})
\begin{equation}
\tilde{H}_n(x) = \frac{ e^{ -x^2/2 } H_n( x)}{\sqrt{2^n\, n! \sqrt{\pi}}}.
\label{Eq:Htilde}   
\end{equation} 
For each Landau level $n$, the eigenvalue given by Eq.~(\ref{Eq:EigEner}) is $k_y-$independent and the energy level is degenerate. The degeneracy $N_\text{L}$ can be obtained by making use  of periodic boundary conditions and requiring  the equilibrium position $x^*$ to be inside the sample \cite{VignaleBook},
 \begin{equation}
 N_\text{L} = \frac{\mathcal{A}}{2\pi \ell_\text{B}^2} = \frac{\Phi}{\Phi_0},
 \end{equation}
 where $\mathcal{A}=L_x \times L_y$ is the sample area, $\Phi = B \mathcal{A}$ is the magnetic flux, and $\Phi_0 $ is the flux quantum.  
 

\section{Theory}\label{sec:theory}
To study the Coulomb interaction between the electron and hole components of an exciton, we use the free-particle wavefunction form in Eq.~(\ref{Eq:LandauWave}) as a basis to build matrix elements of the Coulomb interaction.  We will then use these matrix elements to construct the exciton Hamiltonian in magnetic field.

\subsection{The Coulomb matrix element} \label{sec:CoulombEle}
Second quantization is a convenient formalism to derive matrix elements of the Coulomb  interaction between two charge particles. The second quantization form of the wavefunction in Eq.~(\ref{Eq:LandauWave}) is  (Appendix \ref{ap:p_wave})
\begin{equation} 
| n,k_y \rangle  = (-i)^n\sqrt{  \frac{    2 \pi \ell_\text{B} }{ L_x  }}  \sum_{k_x}    e^{ \mp ik_x k_y \ell^2_\text{B}}   \tilde{H}_n(k_x \ell_\text{B}) \,\,  c^\dagger_{{\bf k}}  |0\rangle .\,\,
\label{Eq:MagWaveF}
\end{equation}
$c^\dagger_{\bf k}$ is the creation operator of a charge particle with wavevector ${\bf k} = (k_x,k_y)$ acting on the vacuum state $| 0 \rangle$ (empty conduction band and full valence band). 

Next, we consider the interaction between two distinguishable particles, represented by creation operators $c^\dagger$ and  $d^\dagger$. Namely, two particles that belong to different bands, and/or reside in different valleys, and/or have opposite spins. The initial state of the two particles is denoted by their quantum numbers $| n_c,k; n_d,p \rangle$  where $\{n_c,n_d\}$ are their LLs and $\{ k,p\}$ are their wavenumbers along the $y$ direction. Hereafter, we omit the $y$ index of wavenumbers $k$ and $p$ to simplify the notation (i.e., $k=k_y$ and $p=p_y$). The omission signifies that the gauge choice in Eq.~(\ref{Eq:LandauGauge}) is such that $[H_0,p_y] = 0$ while $[H_0,p_x] \neq 0$ in Eq.~(\ref{eq:H}), and therefore,  only the $y$ components are good quantum numbers. Using Eq.~(\ref{Eq:MagWaveF}), the Coulomb matrix element can then be written as (Appendix \ref{app:MaDir}) 
\begin{eqnarray}
V_{n_c,k;n_d,p }^{n'_c,k';n'_d,p' } &=& \langle n'_c,k'; n'_d,p'| \hat{V}| n_c,k; n_d,p \rangle  \nonumber \\
&=&\!\sum_{\bf q} e^{i(\pm k \mp p)q_x \ell_\text{B}^2} \,\, W_{n_c;n_d }^{n'_c;n'_d }({\bf q}) \delta_{k;p}^{k';p'}\!(q_y) , \qquad
\label{Eq:MatrixC}
\end{eqnarray}
where quantum numbers of particles in their final states are denoted with prime symbols. ${\bf q} = (q_x, q_y)$ is the transferred wavevector between the particles, and $ \delta_{k;p}^{k';p'}(q_y) \equiv \delta_{k',k+q_y} \delta_{p',p-q_y} $ stems from momentum conservation. The magnetic phase factor has  contributions from the first and second particles via $ e^{\pm i kq_x \ell_\text{B}^2}$ and $ e^{\mp i p q_x \ell_\text{B}^2}$, respectively. The origin of the phase factors is traced back to their different equilibrium positions, $x^*$ (see Appendices \ref{ap:p_wave} and \ref{app:MaDir}). Finally, the interaction term in Eq.~(\ref{Eq:MatrixC}) is 
\begin{equation}
W_{n_c;n_d }^{n'_c;n'_d }({\bf q})  = V({\bf q}) \,\, S_{n_c}^{n'_c}({\bf q}) \,\, S_{n_d}^{n'_d}({- \bf q}), 
\label{Eq:Wq}
\end{equation}
where $V({\bf q})$ is the Fourier transform of the Coulomb potential $V({\bf r})$. The form factor functions $S_{n_c}^{n'_c}({\bf q})$ and $S_{n_d}^{n'_d}({- \bf q})$ reflect momentum conservation where one particle receives $\bf q$ and the other $-{\bf q}$.  This function reads
\begin{equation}
S_{n}^{n'}({\bf q})  = i^{|\Delta n|}\,\, e^{ \pm \frac{i}{2} q_x q_y \ell_\text{B}\pm i \Delta n \theta} \,\, \tilde{L}^{|\Delta n|}_{\tilde{n}} \left( \frac{q^2 \ell_\text{B}^2}{2} \right),
\label{Eq:Sq}
\end{equation}
where $\Delta n  = n' - n$ and $\tilde{n}= \text{min}\!\left\{n,n' \right\}$  are the difference and minimum value of the two LL indices, respectively. $\theta =\arctan\left( q_y/q_x\right)  $ is the angle between ${\bf q}$ and the $x$-axis, and the functions
\begin{equation}
\tilde{L}^m_n(u) = \sqrt{  \frac{ n! }{    \left( n+ m \right)! } \,\,\,  u^{m} \,\,  e^{-u  }  }       \,\,\,\,     L^{m}_{n}\left( u\right)  \label{Eq:Lmntilde}
\end{equation}
are related to the generalized Laguerre polynomials $L^m_n(u)$.  The features and recurrence relations of $\tilde{L}^m_n(u)$ play important role in the numerical calculations (Appendix  \ref{app:LFunc}). Results similar  to the matrix element in Eq.~(\ref{Eq:MatrixC}) have been obtained in Refs.~\cite{Efimkin_PRB2018, Smolenski_PRL2019,MacDonald1994} but for the symmetric gauge. Similar formulas in graphene were developed in Refs.~\cite{Cote_PRB08,Zhang_PRB07,Zhang_PRB08,Goerbig_PRB06}. Equation ~(\ref{Eq:MatrixC}) for  intra-LL scattering between two electrons, $n'_\text{c(d)}=n_\text{c(d)}$, were derived by Goerbig {\it et al.} \cite{Goerbig_PRB04}. 
 
For completeness, we also present the matrix element of the Coulomb interaction between indistinguishable particles. Here, the interaction is between two electrons (or two holes) that belong to the same energy band, reside in the same valley, and have the same spin. There are direct and exchange interaction terms in this case. The direct term is the same as in Eq.~(\ref{Eq:MatrixC}), whereas the exchange term has opposite sign and the particle indices are switched in the final state (Appendix \ref{app:MaExc})
 \begin{eqnarray}
&\,& \langle n_{1}',k_1'; n_{2}',k_2' | \hat{V}| n_{1},k_1; n_{2},k_2 \rangle  \nonumber \\ &\,& \qquad \qquad = \,\,  V_{n_1, k_1;n_2,k_2 }^{n'_1,k'_1; n'_2,k'_2 }  - V_{n_1, k_1;n_2,k_2}^{n'_2,k'_2; n'_1,k'_1 }.  
 \end{eqnarray}

\subsection{The exciton Hamiltonian in magnetic field}
The state of a free electron-hole pair without the Coulomb interaction is
\begin{equation}
|\Psi_{\text{free}} \rangle =   \left| n_\text{e},k_\text{e}; n_\text{h}, k_\text{h}\right\rangle = \left| n_\text{e},k_\text{e}; n_\text{h},K- k_\text{e}\right\rangle, \label{eq:phi_f}
\end{equation}
where $K = k_\text{e} + k_\text{h}$ is the CoM wavevector of the pair along the $y$ direction. $|\Psi_{\text{free}} \rangle$ is eigenstate of the free-pair Hamiltonian,
\begin{equation}
\left(H_0^\text{e} + H_0^\text{h}  \right)  |\Psi_{\text{free}} \rangle =  ( \varepsilon_{n_\text{e}} + \varepsilon_{n_\text{h}} )  |\Psi_{\text{free}} \rangle,  \label{eq:E_f}
\end{equation}
where $H_0^\text{e}$ and $H_0^\text{h}$ are evaluated with the electron and hole effective mass in Eq.~(\ref{eq:H}), respectively.   Similarly, the energies $\varepsilon_{n_\text{e}}$ and $\varepsilon_{n_\text{h}}$ are evaluated through Eq.~(\ref{Eq:EigEner})  with the respective cyclotron frequencies, $\omega_\text{e} = eB / m_\text{e}$ and  $\omega_\text{h} = eB / m_\text{h}$. 

Turning on the Coulomb attraction between the electron and hole, the bound exciton state can be expressed as a superposition of free electron-hole pair states 
\begin{equation}
|\Psi_{K,\text{b}} \rangle = \sum_{n_\text{e},n_\text{h}, k_\text{e}}  \phi^{K}_{n_\text{e},n_\text{h}}\!\!\left( k_\text{e}\right)\,\,\,  \left| n_\text{e},k_\text{e}; n_\text{h},K- k_\text{e}\right\rangle, \label{eq:phi_b}
\end{equation}
where the envelope function $ \phi^{K}_{n_\text{e},n_\text{h}}\!\!\left( k_\text{e}\right)$ is obtained from
\begin{equation}
\left( H_0^\text{e} + H_0^\text{h} + V\left({\bf r}_{\text{e}} -  {\bf r}_{\text{h}} \right) \right) |\Psi_{K,\text{b}} \rangle = E_K |\Psi_{K, \text{b}} \rangle.
\label{Eq:Secular1}
\end{equation}
The formation of the bound exciton state through the electron-hole Coulomb interaction, $V({\bf r}_{\text{e}} -  {\bf r}_{\text{h}} )$, is such that the stronger the attraction between the electron and hole is, more LLs are incorporated in the superposition of Eq.~(\ref{eq:phi_b}). Namely, $ \phi^{K}_{n_\text{e},n_\text{h}}\!\!\left( k_\text{e}\right)$ remains sizable for more pairs of $n_\text{e}$ and $n_\text{h}$.  

We emphasize that $K$ is a constant of motion rather than a variable in Eqs.~(\ref{eq:phi_b})-(\ref{Eq:Secular1}). The reason is that the Coulomb interaction between the electron and hole conserves their center-of-mass wavevector. One interesting consequence  is that the separation between equilibrium positions of the electron and hole  in magnetic field, as given by Eq.~(\ref{Eq:equiPos}),  is also a constant of motion 
\begin{equation}
x^*_\text{e}- x^*_\text{h} = (k_\text{e} + k_\text{h})\ell_\text{B}^2 =  K \ell_\text{B}^2\,. 
\label{Eq:xConser}
\end{equation} 
The result together with the fact that $ \langle {y}_\text{e}  \rangle =   \langle {y}_\text{h}  \rangle =0$ and  $ \langle {x}_\text{e}   - {x}_\text{h}  \rangle = x^*_\text{e}- x^*_\text{h}$  (obtained from  Eqs.~(\ref{Eq:LandauWave}) and (\ref{eq:phi_b}))  leads to the interesting equation
\begin{equation} 
 \langle {\bf r} \rangle = \langle {\bf r}_\text{e} -   {\bf r}_\text{h}   \rangle = \hat{\bf z} \times {\bf K}\,\,   \ell_\text{B}^2
 \end{equation}
 obtained by Lerner and Lozovik in the limit of high magnetic field  \cite{Lerner_JETP80,Lozovik_PRB02}. The equation indicates that the exciton is polarized along the direction perpendicular to the CoM momentum $\bf K$. The result was explained as the distortion of the  wavefunction of the exciton under  effects of opposite Lorentz forces acting on opposite charges of its electron and hole components which move together with the CoM momentum $\bf K$ \cite{Lozovik_PRB02}.  
For an exciton in the light cone ($K \rightarrow 0$),  there is no  distortion and the electron and hole have the same equilibrium position.  

To build a secular matrix equation, we project Eq.~(\ref{Eq:Secular1}) on the state $\left| n_\text{e}',k_\text{e}'; n_\text{h}',K- k_\text{e}' \right\rangle$. After some straightforward algebra, we get that (Appendix~\ref{app:Hamil})
\begin{equation}
 \sum_{n_\text{e},n_\text{h}} \left( T_{n_\text{e},n_\text{h} }^{n'_\text{e},n'_\text{h} } + \widetilde{V}_{n_\text{e},n_\text{h} }^{n'_\text{e},n'_\text{h} } \right) \,\,\,  \phi^{K}_{n_\text{e},n_\text{h}}   = E_K  \,\,\,  \phi^{K}_{n'_\text{e},n'_\text{h}}. 
  \label{Eq:HamilMain}
\end{equation}
 $\phi^{K}_{n_\text{e},n_\text{h}}$ is the contracted envelope function,
\begin{equation}
\phi^{K}_{n_\text{e},n_\text{h}} = \sum_{k_\text{e}} \phi^{K}_{n_\text{e},n_\text{h}}\!\!\left( k_\text{e}\right),
\end{equation} 
$T_{n_\text{e},n_\text{h} }^{n'_\text{e},n'_\text{h} }$ are the kinetic energy matrix elements 
 \begin{eqnarray}
T_{n_\text{e},n_\text{h} }^{n'_\text{e},n'_\text{h} }  &=&    \left[\varepsilon_{n'_\text{e}} + \varepsilon_{n'_\text{h}} \right] \delta_{n_\text{e},n'_\text{e}} \delta_{n_\text{h},n'_\text{h}} \,\,,  \nonumber \\
\label{Eq:Tt}
\end{eqnarray}
and $\widetilde{V}_{n_\text{e},n_\text{h} }^{n'_\text{e},n'_\text{h} }$ are the potential energy matrix elements 
 \begin{eqnarray}
\widetilde{V}_{n_\text{e},n_\text{h} }^{n'_\text{e},n'_\text{h} } &=&  \sum_{\bf q} e^{iKq_x \ell_\text{B}^2} \,\, W_{n_\text{e};n_\text{h} }^{n'_\text{e};n'_\text{h} }({\bf q}).  
\label{Eq:Wt}
\end{eqnarray}
Using Eqs.~(\ref{Eq:Wq})-(\ref{Eq:Lmntilde}), these matrix elements can be written explicitly as (Appendix~\ref{app:Hamil}), 
 \begin{eqnarray}
&\,& \widetilde{V}_{n_\text{e},n_\text{h} }^{n'_\text{e},n'_\text{h} } = - \frac{e^2}{\ell_B} \times  i^{|\Delta n_\text{e}| + \Delta n_\text{e} - |\Delta n_\text{h}| - \Delta n_\text{h}  } \times \nonumber \\ && \quad \int_0^\infty \!\!\! du \, \frac{J_{m}(K\ell_Bu)}{\epsilon(u)} \, \tilde{L}^{|\Delta n_\text{e}|}_{\tilde{n}_\text{e}} \left( \frac{u^2}{2} \right)   \,  \tilde{L}^{|\Delta n_\text{h}|}_{\tilde{n}_\text{h}} \left( \frac{u^2}{2} \right), \,\,\,\,\,\,\,\,\,\label{Eq:Wt1}
\end{eqnarray}
where $\Delta n_{\text{e(h)}}  = n_{\text{e(h)}}' - n_{\text{e(h)}}$, $\tilde{n}_{\text{e(h)}}= \text{min} \{n_{\text{e(h)}},n_{\text{e(h)}}' \}$, $J_m(K\ell_Bu)$ is the $m$-th Bessel function of the first kind, where $m=\Delta n_{\text{e}} - \Delta n_{\text{h}}$, and $\epsilon(u)$ is the dielectric screening function. Unless stated otherwise, we employ the Rytova-Keldysh dielectric model $\epsilon(u) = \epsilon_v(1 + r_0 u/\ell_B)$, where $\epsilon_v$ is the effective dielectric constant of the environment around the 2D sample, and  $r_0 = 2 \pi \alpha /\epsilon_v$ is related to the polarizability $\alpha$ of the sample  \cite{Rytova_MSU67,Keldysh_JETP79,Cudazzo_PRB2011}.

Equations~(\ref{Eq:HamilMain})-(\ref{Eq:Wt1}) are general and can be used to model magneto-excitons in any material with quadratic energy dispersions in the edges of the conduction and valence bands. Equation~(\ref{Eq:HamilMain}) is solved by matrix diagonalization, yielding the energies of the bound exciton states, where the element $ \phi^{K}_{n_\text{e},n_\text{h}} $ in the corresponding eigenvectors represents the weight of the LLs pair  $\{n_\text{e}, n_\text{h}\}$ in these states. Note that this model only calculates the binding energies. It does not deal with the rigid shifts of the conduction and valence energy bands, as these rigid Zeeman energy shifts do not modify the exciton wave functions \cite{Katsch_PRB20}. Also not considered is the Berry curvature flux, which in certain materials is predicted to split the energy of exciton states with opposite angular momentum even at zero magnetic field \cite{Ajit_PRL15,Zhou_PRL15,Trushin_PRL18,Yong_NatMater19}. 

\begin{table*}
\caption{\label{tab:table1} Energies of the ground state ($1s$) and eight lowest excited states of magneto-excitons in hBN-encapsulated WSe$_2$ monolayer, calculated at two different magnetic fields by using the Rytova-Keldysh dielectric screening model  \cite{Rytova_MSU67,Keldysh_JETP79}.  Parameters for the calculations can be found in Appendix \ref{App:Para}.}
\begin{ruledtabular}
\begin{tabular}{c|ccccccccc}
 Magnetic Field  &\multicolumn{9}{c}{Energy~(meV)}\\
  &1$s$& 2$p^-$  & 2$p^+$ & 2$s$ & 3$d^-$ & 3$d^+$ & 3$p^-$ & 3$p^+$ & 3$s$ \\ \hline
 B=10 T&$- 170$ & $ - 53.1$  &$- 52.3 $ & $-39.9 $ & $-22.2$  & $-20.6$ & $ - 19.2$ & $ - 18.4$ & $ - 14.3$  \\
 B=50 T&$- 170$ & $ - 47.2$  &$- 43.3 $ & $-25.9 $ & $-4.6 $     & $3.2$ & $ 10.6$ & $ 14.5$ & $ 22.6$   \\
\end{tabular}
\end{ruledtabular}
\end{table*} 

\subsection{Selection rule of an exciton in the light-cone}\label{sec:SeRule}

The presence of the oscillating Bessel function in Eq.~(\ref{Eq:Wt1}), $J_{m}(K\ell_Bu)$ where $m=\Delta n_\text{e} - \Delta n_\text{h}$, weakens the binding energy between the electron and hole. The oscillatory behavior is worsened when $K$ increases, reflecting the energy increase due to the translational motion of the exciton (i.e., its center-of-mass motion).  The weaker binding can also be understood from Eq.~(\ref{Eq:xConser}), which states that the separation between the equilibrium positions of the electron and hole is commensurate with $K$. That is, $J_{m}(K\ell_Bu) = J_{m}\left((x^*_\text{e}- x^*_\text{h})u/\ell_B\right)$. The binding energy is strongest when the exciton resides in the light cone, $K = 0$, in which case the oscillatory behavior is completely quenched and we get that $J_{m}(0)=1$ when $m=0$ ($\Delta n_\text{e} = \Delta n_\text{h}$), and $J_{m}(0)=0$ when $m\neq 0$ ($\Delta n_\text{e} \neq \Delta n_\text{h}$). Namely, the scattering selection rule of an exciton in the light cone obeys
\begin{equation}
\Delta n = n'_\text{e} - n_\text{e} = n'_\text{h} - n_\text{h}.
\label{Eq:ScatteringSR}
\end{equation}
We continue the analysis for excitons in the light cone, showing how this selection rule limits the way in which LLs of the electrons and holes are paired.  Excitons with finite center-of-mass momentum (i.e., $K\neq 0$), will be analyzed in the next section.

\subsection{Pairing law of electron and hole Landau levels} \label{Result:LLDis}

\begin{figure}
\includegraphics[width=\columnwidth]{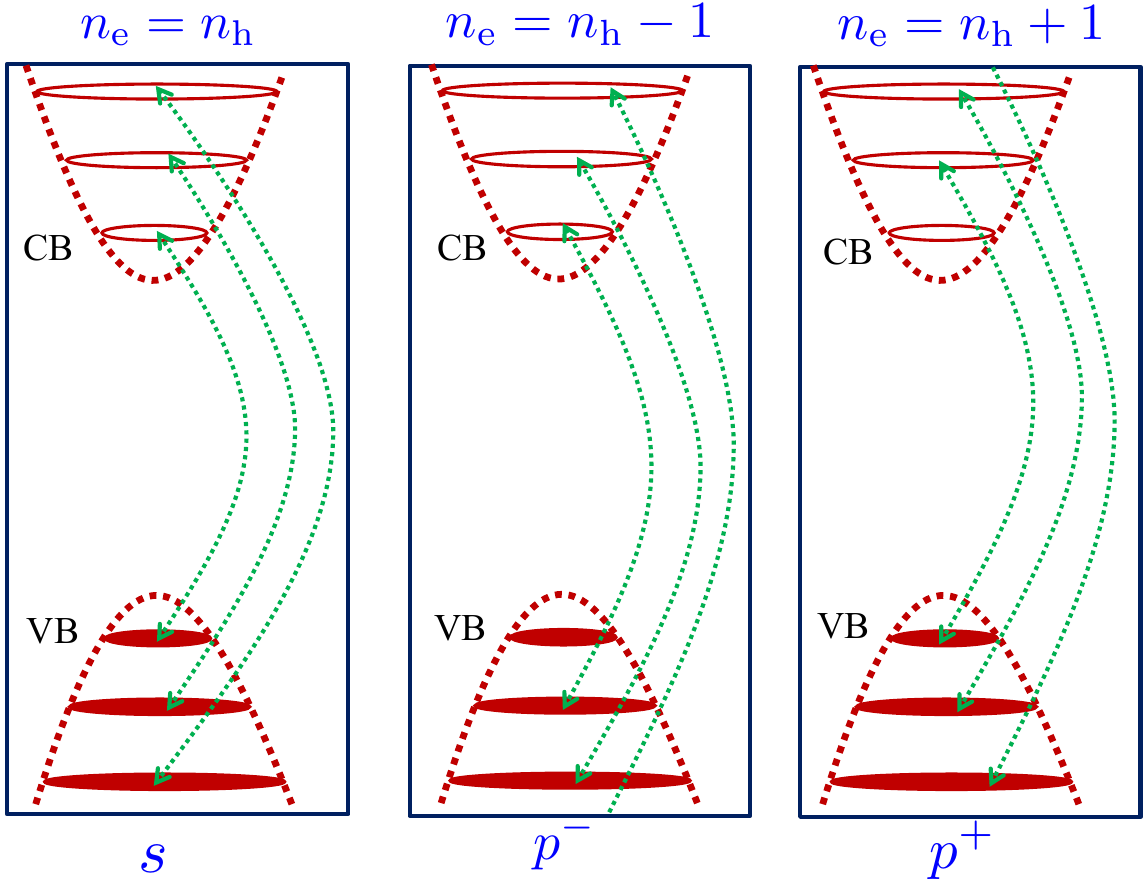}
\caption{ Diagrams of the composition of  $\{s, p^-,p^+\}$ states, showing the corresponding pairing of Landau levels. }\label{fig:LLCom} 
\end{figure}

If we define the difference in LL indices of the electron and hole components, $l = n_\text{e} - n_\text{h}$,  Eq.~(\ref{Eq:ScatteringSR}) indicates that $l$ is another constant of motion,
\begin{equation}
l = l',
\end{equation}
where $l' = n'_\text{e} - n'_\text{h}$.  The potential matrix element vanishes  for scattering between states with different $l$, i.e., $\widetilde{V}_{n+l,n }^{n'+l',n' } =0$ for $l \neq l'$. Consequently,  the Hamiltonian matrix can be partitioned into a block diagonal form. Each block of the partitioned matrix comprises all pairs with $\{n_\text{e}=n+l, n_\text{h}=n\}$, characterized by an integer index $l$. As we will clarify in  Sec. \ref{Sec:Hr}, $l$ is the magnetic quantum number, where diagonalizing the block of pairs with $l=0$ yields $s$-states ($1s$, $2s$, $3s$...). Similarly, diagonalizing the blocks of pairs with $l=\pm1$ yield $p$ states ($2p^\pm$, $3p^\pm$, $4p^\pm$...), and so on (see Fig.~\ref{fig:LLCom}). All in all, the energy spectrum of the 2D magneto-excitons resembles that of a 2D hydrogen model, where states are denoted by their principal  and magnetic quantum numbers (i.e., 1$s$, 2$s$, 2$p^\pm$, ...)\cite{Yang_PRA91}. States with $l = \{ 0,\pm 1, \pm 2, \pm 3,...\}$ assume the conventional respective labeling $\{ s, p^\pm, d^\pm, f^\pm,...\}$.

Putting these pieces together, the secular matrix equation for states with magnetic quantum number $l$ reads
\begin{equation}
 \sum_{n} \left( T_{n,n'}^{l} + \widetilde{V}_{n, n'}^{l} \right)   \phi^{l}_{n}   = E_l   \phi^{l}_{n'}, 
  \label{Eq:HamilMain2}
\end{equation}
where the envelope function is defined by $\phi^{l}_{n} \equiv \ \phi^{K=0}_{n+l,n}$. The kinetic energy matrix elements are, 
 \begin{eqnarray}
T_{n,n'}^{l}  =    \left[ \hbar \omega_\text{e} \left( n+ l+ \frac{1}{2} \right) +  \hbar \omega_\text{h} \left( n + \frac{1}{2} \right) \right] \delta_{n,n'} ,\,\,\,\,\,\,\, \label{Eq:Ekin}
\end{eqnarray}
and the potential energy matrix elements are
\begin{eqnarray}
\widetilde{V}_{n, n'}^{l} = - \frac{e^2}{\ell_B}  \int_0^\infty \!\! \frac{du}{\epsilon(u)} \tilde{L}^{|\Delta n|}_{\tilde{n}+l}  \left( \frac{u^2}{2}  \right)    \tilde{L}^{|\Delta n|}_{\tilde{n}} \left( \frac{u^2}{2}  \right)  , \,\,\,\,\,\,\,\,\,\label{Eq:Vpot}
\end{eqnarray}
where $\Delta n  = n' - n$ and  $\tilde{n}= \text{min}\{n,n' \}$.

\section{RESULTS AND DISCUSSIONS} \label{sec:results}

We have identified that a bound exciton with magnetic quantum number $l$ is constructed by a superposition of free electron-hole pair states, where the electron LL in each pair has $l$ more nodes than in the hole LL. To understand the physical origin of the pairing law, 
\begin{equation}
n_\text{e} = n_\text{h} + l,
\label{Eq:LLCoupling}
\end{equation}     
we recall that the magnetic length $\ell_\text{B}$ and equilibrium positions $x^*_\text{e/h}$ of the electron and hole are identical when the exciton resides in the light cone (Eq.~(\ref{Eq:xConser})). Consequently, the electron and hole components of an $s$-state exciton, wherein $n_\text{e} = n_\text{h}$, have completely overlapping wavefunctions, and therefore, these exciton states bear no magnetic moment.  A nonzero magnetic quantum number $l$ entails a mismatch between the electron and hole wavefunctions (one is more extended than the other; see Appendix \ref{ap:p_wave}), which in turn gives rise to a finite magnetic moment.  Appendix \ref{App:3DExt} shows that the pairing law still holds in quasi-2D and 3D systems in which the  exciton basis  functions and the Coulomb matrix elements are modified by factors accounting for motions of the electron and hole components along the $z$ direction.    In addition, it is worth noting that the pairing law is protected by the energy spacing between exciton states rather than the Landau level spacing. That is, the pairing law still holds even for weak magnetic field as long as the scattering from external disorder sources is not strong enough to mix nearby exciton states (e.g., $1s$ with $2p^{\pm}$).

In the following, we solve Eq.~(\ref{Eq:HamilMain2}) for the $s$, $p$, $d$ and $f$ states of magneto-excitons in the light cone of hBN-encapsulated WSe$_2$ monolayer. Appendix \ref{App:Para} provides information about all quantities and parameters used in the calculations. While we use WSe$_2$ monolayer as a prototype, we  emphasize that the method is general  and that the parameters we use are not key to understand the physics. For example, the theory can be used to model magneto-excitons in  III-V semiconductor quantum wells or one can employ various dielectric functions in Eq.~(\ref{Eq:Vpot}), such as the 3$\chi$ or slab models \cite{VanTuan_PRB18, Cho_PRB18}. Important conclusions we present below are not affected by these choices.  Here, we assume a single-band model for the electron and hole components of the exciton in WSe$_2$. This assumption is applicable in materials whose conduction and valence bands are relatively far from other energy bands (compared with the exciton binding energy). 

 \begin{figure}
\includegraphics[width=\columnwidth]{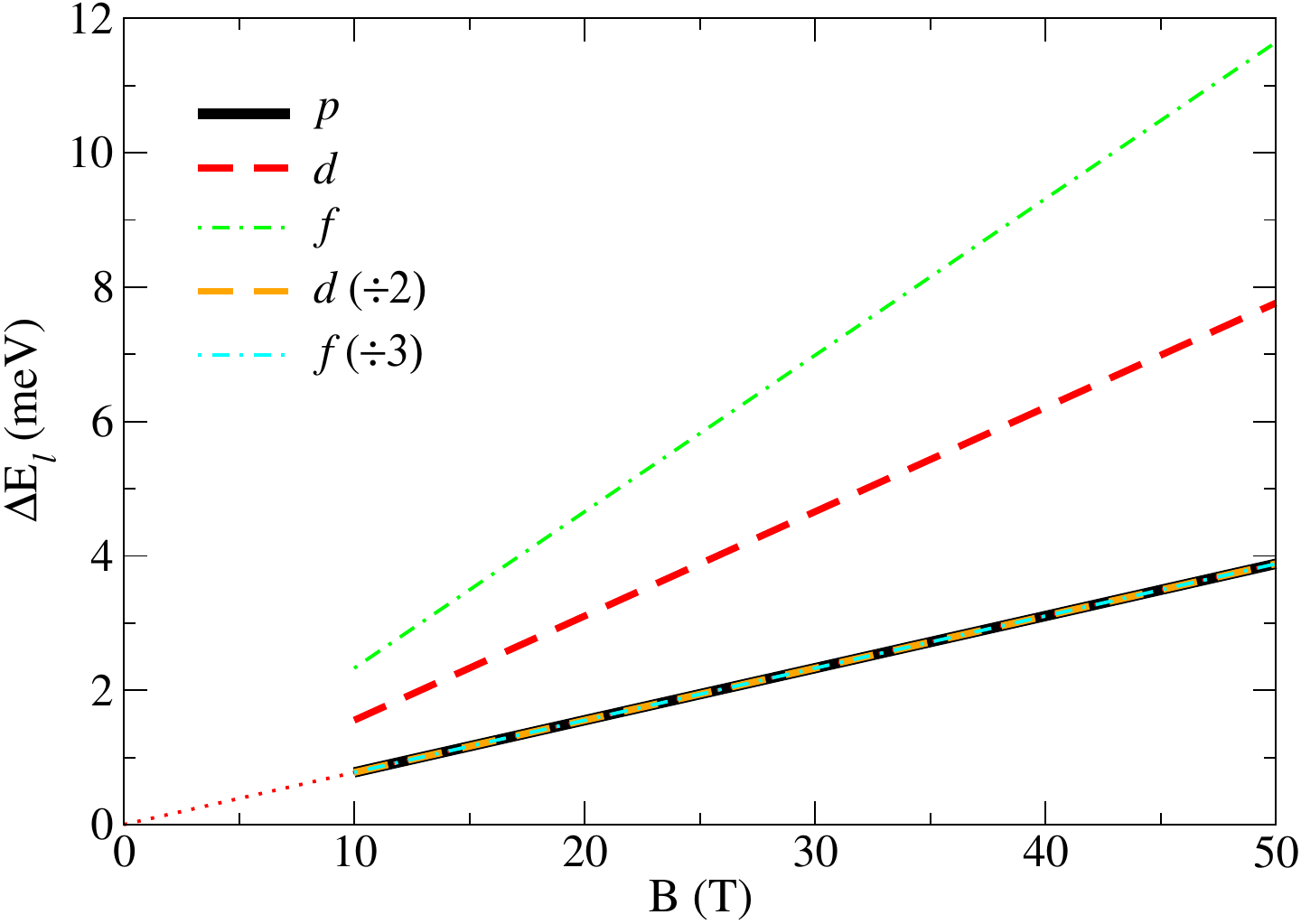}
\caption{ $\Delta E_l =  E_{+l}  -  E_{-l}$ of states with nonzero magnetic quantum number, $\l \ne 0$.   }\label{fig:DEpm} 
\end{figure}

\subsection{Energy splittings}

Table~\ref{tab:table1} shows the exciton energies obtained from diagonalizing the matrix equation, Eq.~(\ref{Eq:HamilMain2}),  when the magnetic fields are $B=10$ and 50$\,$T.  The corresponding exciton energies at zero field, obtained by  the stochastic variational method (SVM) \cite{Varga_CPC_2008,VanTuan_PRB22, VanTuan_SVM22}, are shown in Appendix \ref{app:HighStates} in which we also show the comparison with other works \cite{Spiridonova_PLA20,Donck_PRB18}. We analyze the energy difference  $\Delta E_l =  E_{+l}  -  E_{-l}$ between states with magnetic quantum numbers of opposite sign. The results are shown in Fig.~\ref{fig:DEpm} for $2p$ (solid line), $3d$ (dashed line), and $4f$ states (dash-dotted line). The magnetic field dependence is linear, and it can be extrapolated through the origin at $B = 0$, as shown by the dotted line. Numerical results of \{$p$, $d$, $f$\} states with higher principal quantum numbers, such as \{$3p$, $4d$, $5f$\},  \{$4p$, $5d$, $6f$\}, and so on, yield exactly the  same $\Delta E_l$ as those of \{$2p$, $3d$, $4f$\}. Furthermore, scaling down $\Delta E_d$ by a factor of 2 (orange line) and $\Delta E_f$ by a factor of 3 (cyan line), coincide exactly with $\Delta E_p$ (black solid line). Namely, 
\begin{equation}
\Delta E_l =  l \Delta E_p. 
\end{equation} 
The numerical results match the analytical expression
\begin{equation}
\Delta E_l  = e\hbar \left( m_\text{e}^{-1} - m_\text{h}^{-1} \right) B l = \hbar\left( \omega_\text{e} - \omega_\text{h} \right) l. \label{eq:magmom}
\end{equation}
This relation is understood from the pairing law and the electron-hole symmetry, wherein the state $-l$ can be obtained from the state $+l$ by the transformation $n_\text{e} \leftrightarrow n_\text{h}$. The effective potential matrix element does not change under such transformation ($\tilde{n} \leftrightarrow \tilde{n}+l$ in Eq.~(\ref{Eq:Vpot})). However, the kinetic (cyclotron) energy changes by

\begin{eqnarray}
\!\!\!\!\!\! \!\!\!\! \Delta E_l &=& \left[ \hbar \omega_\text{e} \left( n + l+ \frac{1}{2} \right) +  \hbar \omega_\text{h} \left( n + \frac{1}{2} \right) \right]   \nonumber \\ &-& \left[ \hbar \omega_\text{e} \left( n + \frac{1}{2} \right) +  \hbar \omega_\text{h} \left( n+ l+ \frac{1}{2} \right)   \right] . 
\label{Eq:DelKin}
\end{eqnarray}

\subsection{Real-space Hamiltonian}\label{Sec:Hr}

 A straightforward understanding of the energy splittings and meaning of the quantum number $l$ can be gained from the real-space Hamiltonian of an exciton in magnetic field (Appendix \ref{App:ExHalMag}),
 \begin{equation}
\hat{H}
  =  \hat{K}_r  \,\,  +\,\,  \frac{\hat{L}_z^2}{2 \mu r^2}\,\, + \,\, V(r)  \,\,+ \,\,  \mu^{\text{ex}}_\text{B} \, B \, \frac{\hat{L}_z}{\hbar} \,\,  +   \,\, \frac{e^2 B^2 r^2}{8 \mu}.   
\label{r-Hamil}
\end{equation}
$\hat{K}_r$ is the radial component of the kinetic energy, given by Eq.~(\ref{Eq:RaKr}), $\hat{L}_z = -i \hbar \partial /\partial \theta$ is the angular momentum operator, and $V(r)$ is the Coulomb interaction between the electron and hole where $r = |{\bf r}_\text{e} - {\bf r}_\text{h}|$ is the distance between the electron and hole. The reduced mass of the exciton is $\mu = \left( m_\text{e}^{-1} + m_\text{h}^{-1} \right)^{-1}$ and its Bohr magneton is $\mu^{\text{ex}}_\text{B} =  e \hbar\left( m_\text{e}^{-1} - m_\text{h}^{-1} \right)/2$. The last two terms in Eq.~(\ref{r-Hamil}) describe the exciton interaction with the magnetic field ${\bf B} = B \hat{\bf z}$, where one term comes from the exciton angular  momentum $\hat{L}_z$, and the other term from the quadratic potential confinement of the exciton. The latter gives rise to the diamagnetic shift \cite{Stier_PRL2018,Molas_PRL19,Wang_NanoLett20,Goryca_NatCom19,Spiridonova_PLA20,Liu_PRB19,Pico_PRB25,Have_PRB19}.  Since the Hamiltonian commutes with the exciton angular momentum operator,  $\left[ \hat{H}, \hat{L}_z\right] = 0 $, the eigenvalue of $\hat{L}_z$ is a constant of motion ($\hbar l$) and the Hamiltonian can be written as
 \begin{equation}
\hat{H}_l
  =  \hat{K}_r  \,\,  +\,\,  \frac{\hbar^2 l^2}{2 \mu r^2}\,\, + \,\, V(r)  \,\,+ \,\,  \mu^{\text{ex}}_\text{B} \, B \, l \,\,  +   \,\, \frac{e^2 B^2 r^2}{8 \mu}.   \label{r-Hamil2}
\end{equation}
The energy splitting between exciton states with $|+l\rangle$ and $|\!-l\rangle$ is straightforward, and it is realized by states that `rotate' in opposite directions.  While $|\!-l\rangle$ generates a downward magnetic-moment vector whose interaction with the magnetic field lowers the energy by $l\mu^{\text{ex}}_\text{B} B$, the exciton with $|$$\,$$+$$\,$$l\rangle$ generates an upward magnetic moment vector that raises the energy by $l\mu^{\text{ex}}_\text{B} B$. This energy splitting, $\Delta E_l = 2 l \mu^{\text{ex}}_\text{B}  B$, matches the splitting in Eqs.~\ref{eq:magmom} and (\ref{Eq:DelKin}), showing the physical meaning of  the quantum number $l$.  The matching also indicates that the Zeeman energy effect comes from kinetic (cyclotron)  energies of the electron and hole components, not potential energy of their mutual interaction.

\begin{figure*}[t] 
\centering
\includegraphics[width=15.5cm]{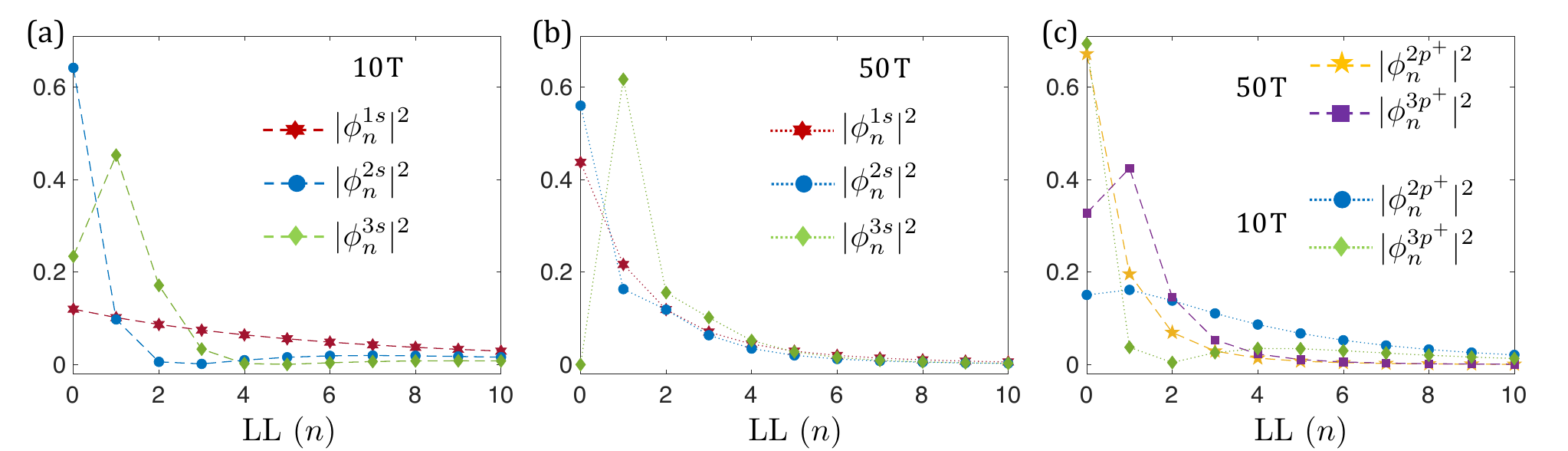}
\caption{ (a) and (b) LL compositions of the 3 lowest energy $s$ states, $\{n_\text{e}=n,\,n_\text{h}=n\}$, when $B\,$=$\,$10 and 50$\,$T, respectively. (c) The same but for the 2 lowest energy $p^+$ states, $\{n_\text{e}=n+1,\,n_\text{h}=n\}$.}\label{fig:LLDis} 
\end{figure*}

\subsection{State composition}
The composition of the $s$ states are shown in Figs.~\ref{fig:LLDis}(a) and (b) for $B = 10$ and 50$\,$T, respectively. While we have used $N=200$ LLs in the calculation, the bound exciton state is mostly made of low-energy LL pairs. Furthermore,  the weight of low energy LLs increases as the magnetic field increases, best seen by comparing the $1s$ states in Figs.~\ref{fig:LLDis}(a) and (b). We also notice that the largest component shifts from $n=0$ to $n=1$ in case of the $3s$ state, attributed to its extended nature.  Figure~\ref{fig:LLDis}(c) shows respective results of the 2$p^+$ and 3$p^+$ states ($l=1$). The corresponding results of $p^-$  are identical but with an exchange of LL indices between the electron and hole, 
\begin{equation}
|\phi^{K=0}_{n_\text{e}=n ,n_\text{h}=n+l}|^2 = |\phi^{K=0}_{n_\text{e}=n+l ,n_\text{h}=n}|^2 .
\end{equation}  
This relation means that one can switch between $p^+$ and $p^-$ ($l=\pm 1$) by exchanging the electron and hole masses or their charge signs. This electron-hole symmetry also applies for states with higher $l$, in accord with Eq.~(\ref{eq:magmom}), and it also means that  states with  opposite  magnetic quantum numbers   ($\pm l$) are degenerate when $m_\text{e} = m_\text{h}$.  

\subsection{Exciton diamagnetic shift}

The diamagnetic shift of the exciton represents its slight repellence by the magnetic field. This effect is seen through the last term in Eq.~(\ref{r-Hamil2}), wherein the magnetic field acts to `squeeze' the electron-hole distance $r$ by introducing a weak confinement potential with quadratic dependence on $B$. In our numerical scheme, we solve Eq.~(\ref{Eq:HamilMain2}) and extract the diamagnetic shift from the average energy of the $\pm l$ states,
\begin{equation}
\Delta E^\text{dia}_l =  \frac{E_{+l}  +  E_{-l}}{2} -  E_{l,0}.
\end{equation}  
$E_{l,0}$ is the state energy at zero magnetic field, where $E_{l,0} = E_{+l}(B=0)  =  E_{-l}(B=0)$. The average energy of the $\pm l$ states cancels out the interaction of the exciton magnetic moment with the magnetic field, thereby helping us to resolve the quadratic field dependence of the diamagnetic shift \cite{Stier_PRL2018,Molas_PRL19,Wang_NanoLett20,Goryca_NatCom19}.

\begin{figure}[h!]
\centering
\includegraphics[width=\columnwidth]{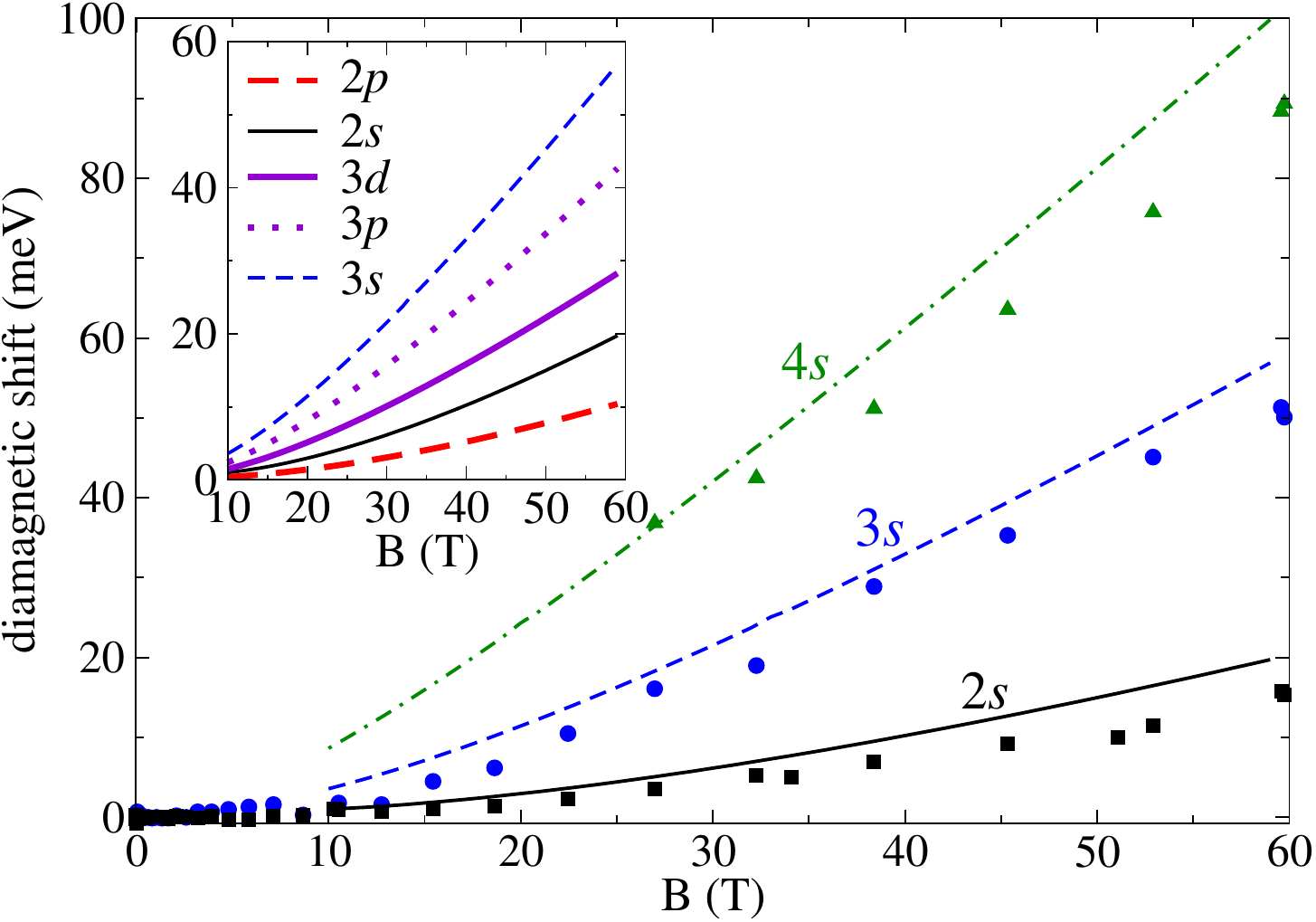}
\caption{Diamagnetic shifts of the $2s$, $3s$, and $4s$ states, where lines denote calculated results and symbols are experimental data from Ref.~\cite{Stier_PRL2018}. Inset: Calculated diamagnetic shifts of the excited states listed in Table \ref{tab:table1}.}\label{fig:Diamag} 
\end{figure}

Figure \ref{fig:Diamag} shows the magnetic field dependence of the diamagnetic shift. The symbols are experimental data taken from Ref.~\cite{Stier_PRL2018}, and the lines are calculated results. Results are shown for the $2s$, $3s$, and $4s$ states and not for $1s$ because the tiny diamagnetic shift of the latter falls within the range of numerical errors. Since the experimental data of the $4s$ state was only identified at $B>27$\,T \cite{Stier_PRL2018}, we have matched its experimental and calculated values at $B = 27$$\,$T instead of at zero field.  The good agreement between experimental and theoretical values reinforces the validity of the model. The maximum deviation between the experiment and theory is 10~meV for the $4s$ state at the highest measured field ($B \simeq 60$$\,$T), which corresponds to $\sim$10\% of the total diamagnetic shift. The theory consistently overestimates the diamagnetic shift, which suggests that a better agreement can be reached by using slightly larger effective masses for the electron and hole. The inset of Fig.~\ref{fig:Diamag} shows the diamagnetic shifts of the $2p$, $3p$, and $3d$ states. Excited states with higher energy (less bound) shift more because of their larger extension, which brings in a stronger confinement potential, evidenced by assigning larger $\langle r^2 \rangle$ in the last term of Eq.~(\ref{r-Hamil2}). 

\subsection{Finite center-of-mass momentum}
\begin{figure}
\includegraphics[width=\columnwidth]{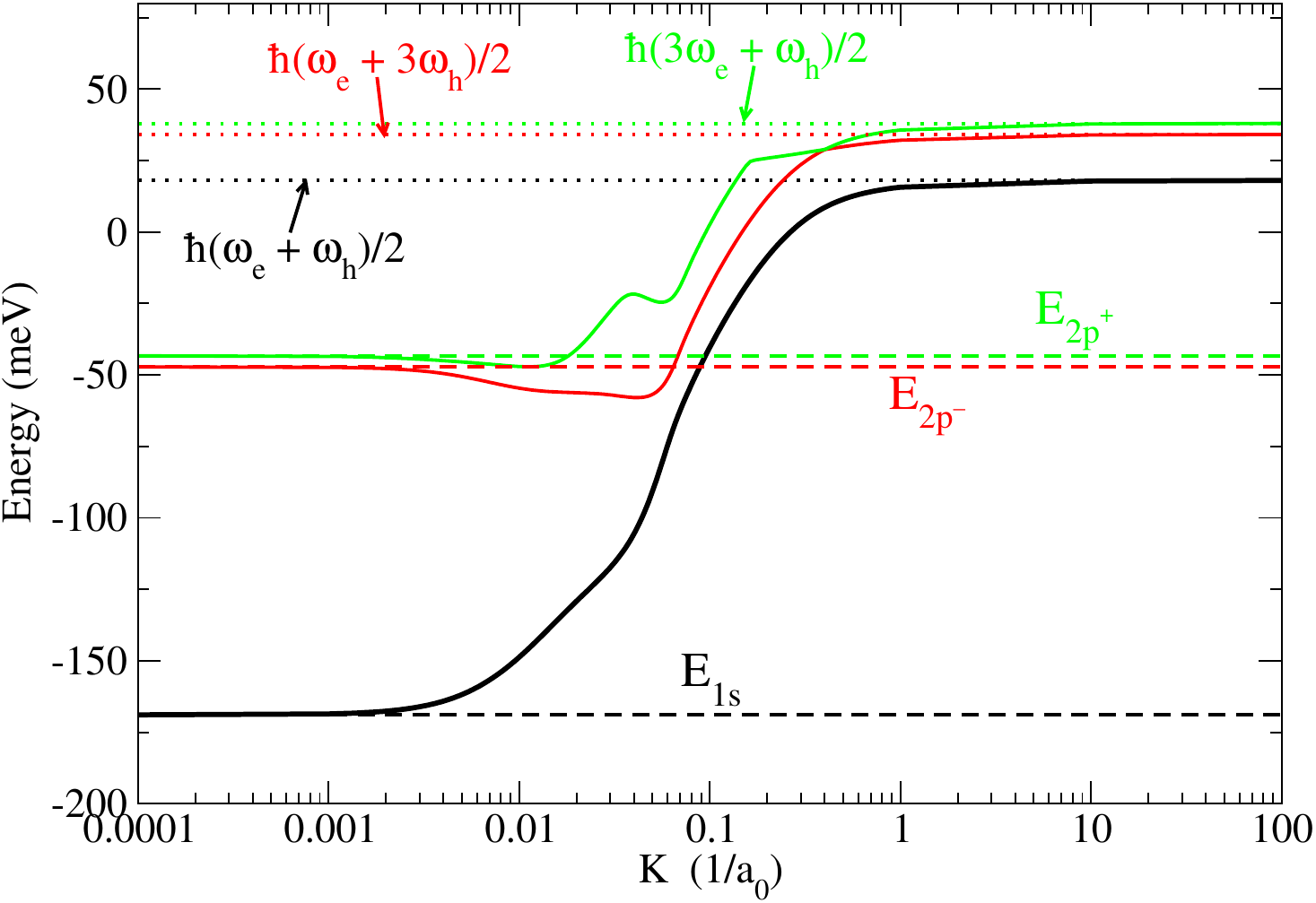}
\caption{Energies of the three lowest energy states as a function of the exciton center-of-mass momentum at $B = 50$~T ($a_0=0.53$~$\AA$ is the Bohr radius in atomic units).}\label{fig:EK} 
\end{figure}

The energy dispersion of magneto-excitons has been considered by Lerner and Lozovik in Refs.~\cite{Lerner_JETP80,Lozovik_PRB02}, who showed a transition from exciton-like states at low CoM momentum  to Landau-level-like states at high momentum. They also showed that the energy dispersion is non-monotonic for all but the ground state. Here, the CoM momentum effects are considered by solving Eq.~(\ref{Eq:HamilMain}) using finite $K$ in Eq.~(\ref{Eq:Wt1}). Unlike the case of  $K=0$, an exciton with finite CoM momentum has preferred direction along which we choose the $y$ axis of the Landau-gauge coordinate.  Therefore, $K$ becomes the total momentum of the CoM and a constant of motion of the system.  Figure \ref{fig:EK} shows the   energy dispersions for excitons in  the three lowest  states, $\{1s, 2p^-,2p^+\}$,  when $B = 50$ T. The results are in good agreement with the ones in Refs.~\cite{Lerner_JETP80,Lozovik_PRB02}; i.e., exciton energies start from their bound-state values of $\{ E_{1s}, E_{2p^-},E_{2p^+}\}$ at low $K$ and approach their corresponding Landau levels $\{ \frac{1}{2}\hbar(\omega_\text{e}+\omega_\text{h}), \frac{1}{2}\hbar(\omega_\text{e}+3\omega_\text{h}), \frac{1}{2}\hbar(3\omega_\text{e}+\omega_\text{h})\}$ at large $K$. The non-monotonic behaviors of the excited states, $p^{\pm}$,  in  Fig.~\ref{fig:EK} are similar to the ones in Ref.~\cite{Lerner_JETP80} which employed a perturbative treatment for the Coulomb potential by considering an extremely strong magnetic field. Here, although we have a strong Coulomb potential, the confluence of $p$ states and large CoM momentum weakens the contribution of the Coulomb potential and validate the perturbative approach. Specifically,  the vanishing wavefunction of  $p$ states  at a short electron-hole distance mitigates the contribution of the short-range Coulomb potential. Furthermore,  the large CoM momentum dictates large separation between equilibrium
positions of the electron and hole in magnetic field (see Eq.(\ref{Eq:xConser})), validating the perturbative contribution of the Coulomb potential. 

\section{Summary and Outlook} \label{sec:con}

We have presented a Hamiltonian model that employs the Landau quantization of electrons and holes to study the spectrum, Zeeman, and diamagnetic energy shifts of magneto-excitons in two-dimensional semiconductors. The model can be used to study the ground and excited states of  magneto-excitons in any material with quadratic energy dispersion. While the model is different than the conventional real-space model of a charge-neutral exciton, wherein the relative motion of the electron and hole is not subjected to Landau quantization, we have shown that the two physical pictures are mutually supportive perspectives of the same problem. 

Using the Hamiltonian of an exciton in the light cone, we have identified a scattering selection rule between the Landau levels of the electron and hole. For an exciton with magnetic quantum number $l$, the scattering selection rule leads to an important pairing law between the electron and hole Landau levels. The pairing law is a central result of this work, which not only greatly simplifies the numerical calculation, but it additionally sheds light on the magnetic moment of the exciton and its interaction with the magnetic field. 

The theory presented in this work can be used to find the effective masses of electrons and holes through the spectrum of magneto-excitons. For example, using nonlinear optics, such as two-photon  excitation spectroscopy \cite{He_PRL14,Ye_Nat14,Wang_PRL15,Zhu_PRL23}, one can probe the exciton states $p^\pm$, whose magnetic-field induced energy splitting can be used to extract the value of $(m_\text{e}^{-1} - m_\text{h}^{-1})$. On the other hand, the exciton reduced mass $\mu = (m_\text{e}^{-1} + m_\text{h}^{-1})^{-1}$ can be extracted from the diamagnetic shift of the $s$-states \cite{Stier_PRL2018,Molas_PRL19,Wang_NanoLett20,Goryca_NatCom19}. Combined together, one can find the effective mass of an electron in the conduction band or of a hole in the valence band, without having to deal with changes to the effective mass that are caused by many-body interactions in electrostatically-doped semiconductors.

Other than magneto-excitons, the Coulomb matrix elements derived in Section \ref{sec:CoulombEle}  are applicable to investigations of other problems. The direct interaction matrix elements can be used to study Fractional Quantum Hall states, while the exchange matrix elements can be used to evaluate the band-gap renormalization of resident carriers in magnetic field. The latter can be used to quantify the enhancement of the spin and valley polarizations when an electrostatically-doped semiconductor is subjected to strong magnetic field  \cite{Liu_PRL20}.

\acknowledgments{We thank Scott Crooker for providing the experimental data in Fig.~\ref{fig:Diamag}. This work is supported by the Department of Energy, Basic Energy Sciences, Division of Materials Sciences and Engineering under Award No. DE-SC0014349.}


\appendix
\section{Real-space and momentum representations of free-particle wavefunctions} \label{ap:p_wave}
\begin{figure}
\centering
\includegraphics[width=6.5cm]{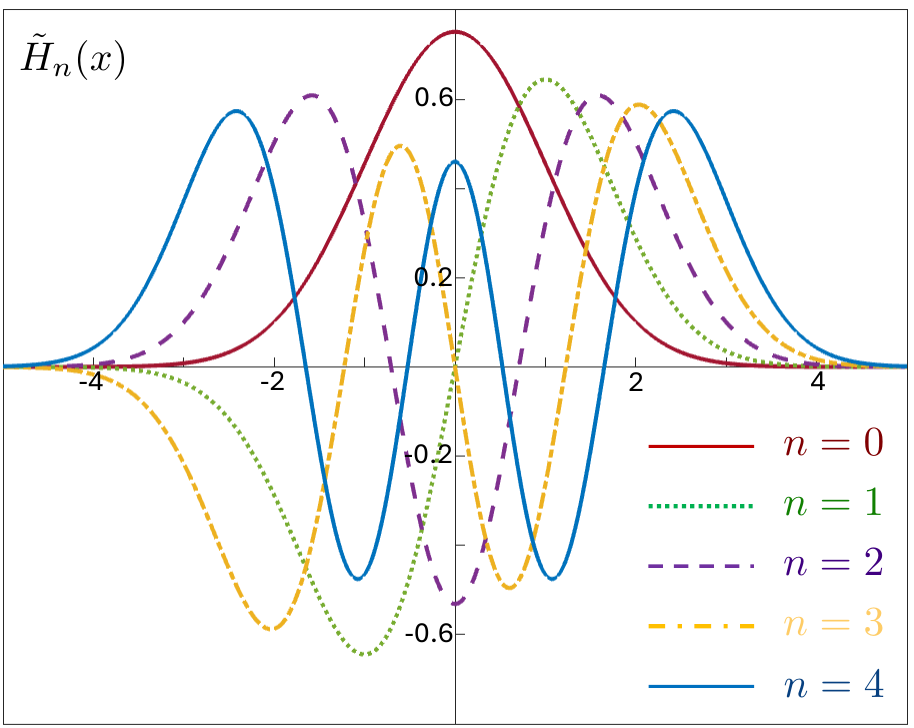}
\caption{ The five lowest  Hermite-Gaussian functions.}\label{fig:Hn} 
\end{figure}

The real-space wavefunction of a charge particle in magnetic field under the Landau gauge is given by Eq.~(\ref{Eq:LandauWave}). The wavefunction is a plane wave in the $y$ direction, modulated by Hermite-Gaussian functions  $\tilde{H}_n(x)$  along the $x$ direction. Following their definition in Eq.~(\ref{Eq:Htilde}), the Hermite-Gaussian functions make an orthonormal basis. As shown in Fig.~\ref{fig:Hn}, $\tilde{H}_n(x)$ are symmetric (antisymmetric) functions for even (odd) $n$, where $n$ is also the number of roots.  

Making use of the following property of Hermite polynomials
\begin{equation}
\frac{1}{\sqrt{2 \pi} } \int_{-\infty}^{\infty} e^{-\frac{1}{2}x^2} H_n(x) e^{-i k x} dx =(-i)^n e^{-\frac{1}{2}k^2} H_n(k),     \nonumber 
\end{equation}
the p-representation of the wavefunction in Eq.~(\ref{Eq:LandauWave}) is obtained from the Fourier transform
\begin{eqnarray} 
 \psi_{n}({\bf k}) &= &  \frac{1}{\sqrt{L_x}} \int_{-\infty }^{\infty} \frac{1}{\sqrt{\ell_\text{B} }}  \,\, \tilde{H}_n\left( \frac{ x \mp k_y \ell_\text{B}^2}{\ell_\text{B}} \right)  e^{-ik_xx} dx 
  \nonumber \\
  &=&  (-i)^n\sqrt{  \frac{    2 \pi \ell_\text{B} }{ L_x  }} \,\,\,    e^{ \mp ik_x k_y \ell^2_\text{B}}\,\,\,   \tilde{H}_n(k_x \ell_\text{B})\,.
  \label{app:Fourier}
\end{eqnarray}
The second-quantization form of the wavefunction is then
\begin{equation} 
| n,k_y \rangle  = i^{-n}\sqrt{  \frac{    2 \pi \ell_\text{B} }{ L_x  }}  \sum_{k_x}    e^{ \mp ik_x k_y \ell^2_\text{B}}   \tilde{H}_n(k_x \ell_\text{B}) \,\,  c^\dagger_{{\bf k}}  |0\rangle . \label{eq:2ndq}
\end{equation}
The phase factor $e^{ \mp ik_x k_y \ell^2_\text{B}} =e^{ - ik_x x^*}  $ comes from the equilibrium position $x^* = \pm k_y \ell_\text{B}^2$. The orthonormality of the wavefunction can be verified from
\begin{eqnarray} 
\!\!\!\!\!\!\!\!\! &\!& \!\!\!\!\!\! \langle m,p_y | n,k_y \rangle  = i^{m-n}  \frac{    2 \pi \ell_\text{B} }{ L_x  }  \sum_{k_x, p_x}    e^{ \mp i(k_x k_y - p_x p_y) \ell^2_\text{B}} \nonumber \\
\!\!\!\!\!\!\!\!\! &\!& \qquad\qquad\qquad \times \tilde{H}^*_m(p_x \ell_\text{B})  \tilde{H}_n(k_x \ell_\text{B}) \,\, \langle 0 | c_{\bf p} c^\dagger_{{\bf k}}  |0\rangle  \nonumber  \\
\!\!\!\!\!\!\!\!\! &\!& \!\!\!\!\!\! = i^{m-n} \delta_{k_y,p_y }  \frac{    2 \pi \ell_\text{B} }{ L_x  }  \sum_{k_x}  \tilde{H}^*_m(k_x \ell_\text{B})  \tilde{H}_n(k_x \ell_\text{B})    \nonumber \\
\!\!\!\!\!\!\!\!\! &\!& \!\!\!\!\!\! = i^{m-n}  \delta_{k_y,p_y }   \int_{-\infty}^{\infty}  \tilde{H}_m(u)  \tilde{H}_n(u) du   = \delta_{n,m} \,\,  \delta_{k_y,p_y }\,.\,\,\,\,
\end{eqnarray}

\section{Matrix element for direct scattering} \label{app:MaDir}
The matrix element in Eq.~(\ref{Eq:MatrixC}) is derived by rewriting the free-particle wavefunction in Eq.~(\ref{eq:2ndq}) as 
\begin{equation} 
| n,k_y \rangle  = i^{-n}\sqrt{  \frac{    2 \pi \ell_\text{B} }{ L_x  }}  \sum_{k_x}    e^{ \mp ik_x k_y \ell^2_\text{B}}   \tilde{H}_n(k_x \ell_\text{B}) \,\,  | {\bf k} \rangle  ,
\end{equation}
where $ |{\bf k} \rangle  = c^\dagger_{{\bf k}}  |0\rangle  $, and the real-space representation of the plane wave is $\langle {\bf r} |   {\bf k} \rangle = e^{i{\bf k.r}}/\sqrt{A}$. The detailed derivation of the direct interaction component of the Coulomb matrix element follows from
\begin{widetext}

\begin{eqnarray}
V_{n_c,k_y; n_d, p_y }^{n'_c,k'_y; n'_d,p'_y } &=& \langle n'_c,k'_y; n'_d,p'_y| \hat{V}| n_c,k_y; n_d,p_y \rangle  \nonumber \\
&=& (i)^{\Delta n_c }  \frac{    2 \pi \ell_\text{B} }{ L_x  }  \sum_{k_x,k'_x}   e^{ \mp i \left(k_x k_y - k'_x k'_y  \right) \ell^2_\text{B}}\,\,\,   \tilde{H}_{n_c}(k_x \ell_\text{B})\,\,\,   \tilde{H}^*_{n'_c}(k'_x \ell_\text{B})    \nonumber \\ 
&\times&  (i)^{\Delta n_d}  \frac{    2 \pi \ell_\text{B} }{ L_x  }  \sum_{  p_x,p'_x}   e^{ \mp i \left(p_x p_y - p'_x p'_y  \right) \ell^2_\text{B}} \,\,\,  \tilde{H}_{n_d}(p_x \ell_\text{B}) \,\,\,  \tilde{H}^*_{n'_d}(p'_x \ell_\text{B}) \,\,\,\,  \langle {\bf k}';{\bf p}' | \hat{V} | {\bf k};{\bf p} \rangle    \nonumber \\
&=& (i)^{\Delta n_c }  \frac{    2 \pi \ell_\text{B} }{ L_x  }  \sum_{k_x,k'_x}   e^{ \mp i \left(k_x k_y - k'_x k'_y  \right) \ell^2_\text{B}}\,\,\,   \tilde{H}_{n_c}(k_x \ell_\text{B})\,\,\,   \tilde{H}^*_{n'_c}(k'_x \ell_\text{B})    \nonumber \\ 
&\times&  (i)^{\Delta n_d}  \frac{    2 \pi \ell_\text{B} }{ L_x  }  \sum_{  p_x,p'_x}   e^{ \mp i \left(p_x p_y - p'_x p'_y  \right) \ell^2_\text{B}} \,\,\,  \tilde{H}_{n_d}(p_x \ell_\text{B}) \,\,\,  \tilde{H}^*_{n'_d}(p'_x \ell_\text{B}) \,\,   \sum_{\bf q} V({\bf q})\,\,\,  \delta_{ k'_x, k_x+q_x} \delta_{ k'_y, k_y + q_y}  \delta_{ p'_x, p_x-q_x} \delta_{ p'_y, p_y - q_y}     \nonumber \\
&=&   \sum_{\bf q} V({\bf q}) \,\, (i)^{\Delta n_c }   \frac{    2 \pi \ell_\text{B} }{ L_x  }   \sum_{k_x}   e^{ \mp i \left[k_x k_y - (k_x+q_x)( k_y+q_y)  \right] \ell^2_\text{B}} \,\,\,  \tilde{H}_{n_c}\left(k_x \ell_\text{B}\right)  \,\,\,  \tilde{H}^*_{n'_c}\left((k_x +q_x) \ell_\text{B}\right)    \nonumber \\ 
&\times& \qquad \qquad   (i)^{\Delta n_d}  \frac{    2 \pi \ell_\text{B} }{ L_x  }   \sum_{p_x}  e^{ \mp i \left[p_x p_y - (p_x - q_x)(p_y -q_y)  \right] \ell^2_\text{B}} \,\,\,  \tilde{H}_{n_d}\left(p_x \ell_\text{B}\right)  \,\,\,  \tilde{H}^*_{n'_d}\left((p_x-q_x) \ell_\text{B}\right) \,\,\,  \delta_{ k'_y, k_y + q_y}  \delta_{ p'_y, p_y - q_y}     \nonumber \\
&=& \sum_{\bf q} V({\bf q})  \,\,     e^{ \pm i  k_y q_x    \ell^2_\text{B}}  \,\,\, \,\,  (i)^{\Delta n_c }\,\, e^{ \pm i  q_x q_y    \ell^2_\text{B}}  \,\,\, \frac{    2 \pi \ell_\text{B} }{ L_x  } \sum_{k_x}   e^{ \pm i k_xq_y   \ell^2_\text{B}} \,\,\,  \tilde{H}_{n_c}\left(k_x \ell_\text{B}\right)  \,\,\,  \tilde{H}^*_{n'_c}\left((k_x +q_x) \ell_\text{B}\right)  \nonumber \\ 
&\times& \qquad \qquad    e^{ \mp i   p_yq_x   \ell^2_\text{B}}  \,\,\,\,\,(i)^{\Delta n_d } \,\,  e^{ \pm i  q_x q_y    \ell^2_\text{B}}    \,\,\, \frac{    2 \pi \ell_\text{B} }{ L_x  } \sum_{p_x}  e^{ \mp i p_x q_y \ell^2_\text{B}} \,\,\,  \tilde{H}_{n_d}\left(p_x \ell_\text{B}\right)  \,\,\,  \tilde{H}^*_{n'_d}\left((p_x-q_x) \ell_\text{B}\right) \,\,\,  \delta_{ k'_y, k_y + q_y}  \delta_{ p'_y, p_y - q_y}     \nonumber \\
&=& \sum_{\bf q} e^{i(\pm k_y \mp p_y)q_x \ell_\text{B}^2} \,\, V({\bf q}) \,\, S_{n_c}^{n'_c}({\bf q}) \,\, S_{n_d}^{n'_d}({- \bf q})  \,\,\, \delta_{k'_y,k_y+q_y} \delta_{p'_y,p_y-q_y}\,,
\label{Eq.VDeri}
\end{eqnarray}
where $\Delta n_{c/d} =n'_{c/d} - n_{c/d} $, ${\bf k}=(k_x,k_y), {{\bf k'} = (k'_x, k'_y)}$, ${\bf p}=(p_x,p_y)$, $ {{\bf p'} = (p'_x, p'_y)}$, and
\begin{eqnarray}
S_{n}^{n'}({\bf q})  &= & (i)^{\Delta n }     \,\,\, e^{ \pm i  q_x q_y    \ell^2_\text{B}}  \,\,\, \frac{    2 \pi \ell_\text{B} }{ L_x  } \sum_{k_x}   e^{ \pm i k_xq_y   \ell^2_\text{B}} \,\,\,  \tilde{H}_{n}\left(k_x \ell_\text{B}\right)  \,\,\,  \tilde{H}^*_{n'}\left((k_x +q_x) \ell_\text{B}\right)   \nonumber \\
 &= & (i)^{\Delta n }     \,\,\, e^{ \pm i  q_x q_y    \ell^2_\text{B}}  \,\,\, \ell_\text{B} \int_{-\infty}^\infty     \tilde{H}_{n}\left(k_x \ell_\text{B}\right)  \,\,\,  \tilde{H}^*_{n'}\left((k_x +q_x) \ell_\text{B}\right)\,\,\, e^{ \pm i k_xq_y   \ell^2_\text{B}} \,\,\,   dk_x   \nonumber \\
 &= & (i)^{\Delta n }     \,\,\, e^{ \pm i  \alpha \beta }  \,\,\,  \int_{-\infty}^\infty    \tilde{H}_{n}\left(x\right)  \,\,\,  \tilde{H}^*_{n'}\left(x +\alpha\right) \,\,\, e^{ \pm i \beta x} \,\,\,   dx   \nonumber \\
 &= & (i)^{\Delta n }     \,\,\, e^{ \pm \frac{i}{2}  \alpha \beta }  \,\,\,  \int_{-\infty}^\infty    \tilde{H}_{n}\left(x-\frac{\alpha}{2}\right)  \,\,\,  \tilde{H}^*_{n'}\left(x +\frac{\alpha}{2}\right) \,\,\, e^{ \pm i \beta  x} \,\,\,   dx   \nonumber \\
 &= &  e^{ \pm \frac{i}{2}  \alpha \beta }  \frac{(i)^{\Delta n }}{\sqrt{\pi \,\, 2^{n+n'} \,\,n! \,\,n'! } }   \int_{-\infty}^\infty   e^{-\frac{1}{2} \left(x-\frac{\alpha}{2}\right)^2}  H_{n}\left(x-\frac{\alpha}{2}\right)  \,\,\, e^{-\frac{1}{2} \left(x+\frac{\alpha}{2}\right)^2} H_{n'}\left(x +\frac{\alpha}{2}\right) \,\,\, e^{ \pm i \beta  x} \,\,\,   dx    \nonumber \\
 &= &  e^{ \pm \frac{i}{2}  \alpha \beta }  \frac{(i)^{\Delta n }}{\sqrt{\pi \,\,2^{n+n'}\,\, n!\,\, n'! } }  \,\, \, T_n^{n'}\left(\alpha,\pm \beta \right).
\label{Eq:STRela}
\end{eqnarray}
The dimensionless parameters are $\alpha = q_x \ell_\text{B}$ and $\beta = q_y \ell_\text{B}$. In addition,
\begin{equation}
T_n^{n'}\left(\alpha,\beta \right) =    \int_{-\infty}^\infty   e^{-\frac{1}{2} \left(x-\frac{\alpha}{2}\right)^2}  H_{n}\left(x-\frac{\alpha}{2}\right)  \,\,\, e^{-\frac{1}{2} \left(x+\frac{\alpha}{2}\right)^2} H_{n'}\left(x +\frac{\alpha}{2}\right) \,\,\, e^{  i \beta  x} \,\,\,   dx,
\end{equation} 
which has the following properties,
\begin{equation}
T_n^{n'}\left(\alpha,\beta \right)  = \left( T^n_{n'}\left(-\alpha,-\beta \right)  \right)^*,
\end{equation}
\begin{equation}
T_n^{n'}\left(\alpha,\beta \right)  = \left( - 1 \right)^{n+n'}\left( T^n_{n'}\left(\alpha,\beta \right)  \right)^*,
\label{Eq:TPro2}
\end{equation}
and the recurrence relation
\begin{equation}
T_n^{n'}\left(\alpha,\beta \right)  = - \left( \alpha + 2i \frac{\partial}{\partial \beta} \right)T_{n-1}^{n'}\left(\alpha,\beta \right)  - 2\left( n-1 \right)T_{n-2}^{n'}\left(\alpha,\beta \right). 
\end{equation}
One can use mathematical induction and the above recurrence relation  to prove the following formula  for $n' \ge n$
\begin{equation}
T_n^{n'}\left(\alpha,\beta \right) = \sqrt{\pi}\,\, 2^n\,\, n!  \,\, \left( \alpha + i \beta   \right)^{n'-n}\,\, e^{-\frac{\alpha^2 + \beta^2}{4}}\,\, L_{n}^{n'-n} \left( \frac{\alpha^2 + \beta^2}{2}\right),
\end{equation}
where $L_{n}^{m}(x)$ are the generalized Laguerre polynomial. Using Eq.~(\ref{Eq:TPro2}), we get the following when $n>n'$ 
\begin{equation}
T_n^{n'}\left(\alpha,\beta \right)= (-1)^{n+n'} \sqrt{\pi}\,\, 2^{n'}\,\, n'! \,\, \left( \alpha - i \beta   \right)^{n-n'}\,\, e^{-\frac{\alpha^2 + \beta^2}{4}}\,\, L_{n'}^{n-n'} \left( \frac{\alpha^2 + \beta^2}{2}\right). 
\end{equation}
Recalling that $\alpha = q_x \ell_\text{B}$,  $\beta = q_y \ell_\text{B}$, and 
\begin{equation}
q_x \pm iq_y = q e^{\pm i\theta}
\end{equation}
where $\theta =\angle({\bf q},\hat{{\bf x}})$, we obtain
\begin{eqnarray}
 T_n^{n'}\left(q_x \ell_\text{B}, \pm  q_y \ell_\text{B} \right) &=&    \sqrt{\pi} \,\, 2^n\,\, n! \,\, \left( q \ell_\text{B}e^{\pm i\theta}   \right)^{\Delta n} \,\, e^{-\frac{1}{4}q^2\ell_\text{B}^2}  \,\, L_{n}^{\Delta n} \left( \frac{q^2 \ell_\text{B}^2}{2} \right) \qquad \mbox{for} \quad \Delta n>0 \\
 T_n^{n'}\left(q_x \ell_\text{B}, \pm  q_y \ell_\text{B} \right) &=&    \sqrt{\pi} \,\, 2^{n'}\,\, n'! \,\, \left( -q \ell_\text{B}e^{\mp i\theta}   \right)^{-\Delta n} \,\, e^{-\frac{1}{4}q^2\ell_\text{B}^2} \,\, L_{n'}^{-\Delta n} \left( \frac{q^2 \ell_\text{B}^2}{2} \right) \qquad \mbox{for} \quad \Delta n<0
\end{eqnarray}
Substituting these expressions into Eq.~(\ref{Eq:STRela}), we obtain a general result for the form factors of both $\Delta n \ge 0$ and $\Delta n < 0$,
\begin{equation}
S_{n}^{n'}({\bf q})  = i^{|\Delta n|}\,\, e^{ \pm \frac{i}{2} q_x q_y \ell_\text{B}\pm i \Delta n \theta} \,\, \tilde{L}^{|\Delta n|}_{\tilde{n}} \left( \frac{q^2 \ell_\text{B}^2}{2} \right).
\label{app:Sq}
\end{equation}
The upper (lower) sign is for an electron (hole), and $\tilde{n}= \text{min} \left\{n,n' \right\}$ is the minimum value between LL indices of the initial ($n$) and final ($n'$) states. The orthonormal functions are
\begin{equation}
\tilde{L}^m_n(x) = \sqrt{  \frac{ n! }{    \left( n+ m \right)! } \,\,\,  x^{m} \,\,  e^{-x  }  }       \,\,\,\,     L^{m}_{n}\left( x\right). 
\end{equation}
Properties  of these functions are detailed in Appendix \ref{app:LFunc}.

When evaluating $S_{n}^{n'}({- \bf q})$, one should note that while $q_x q_y$ is the same for $\bf q$ and $\bf -q$, the angle $\theta$ changes $\pi$ under the transformation, i.e., $\theta_{-{\bf q}} = \theta_{{\bf q}} +\pi$. As a consequence, one obtains
\begin{equation}
S_{n}^{n'}({- \bf q})  = i^{-|\Delta n|}\,\, e^{ \pm \frac{i}{2} q_x q_y \ell_\text{B}\pm i \Delta n \theta} \,\, \tilde{L}^{|\Delta n|}_{\tilde{n}} \left( \frac{q^2 \ell_\text{B}^2}{2} \right).
\label{app:Sq1}
\end{equation}
which is the negative of $S_{n}^{n'}({ \bf q})$ when $|\Delta n|$ is odd.

\section{Matrix element for  exchange scattering} \label{app:MaExc}
To calculate the matrix element of the exchange scattering, it is more convenient to write the free-particle wavefunction in second quantization, i.e., in form of Eq.~(\ref{Eq:MagWaveF}). Because the two interacting particles are identical, there is only one kind of operator. The matrix element is

\begin{eqnarray}
 \langle n'_1,k'_y; n'_2,p'_y| \hat{V}| n_1,k_y; n_2,p_y \rangle  &=& (i)^{\Delta n_1 + \Delta n_2}  \! \left( \frac{    2 \pi \ell_\text{B} }{ L_x  } \right)^2   \!\!\!\! \sum_{k_x,k'_x;  p_x,p'_x}  \!\!\! e^{ \mp i \left(k_x k_y - k'_x k'_y  \right) \ell^2_\text{B}}\,\,\,   \tilde{H}_{n_1}(k_x \ell_\text{B})\,\,\,   \tilde{H}^*_{n'_1}(k'_x \ell_\text{B})    \nonumber \\ 
&\times& \langle 0 ;0 |\,\, c_{{\bf p}'} c_{{\bf k}'} \,\,\, \hat{V} \,\,\, c^\dagger_{\bf k} c^\dagger_{\bf p} |0; 0 \rangle   \,\,\,\,\, \,\,\, e^{ \mp i \left(p_x p_y - p'_x p'_y  \right) \ell^2_\text{B}} \,\,\,  \tilde{H}_{n_2}(p_x \ell_\text{B}) \,\,\,  \tilde{H}^*_{n'_2}(p'_x \ell_\text{B}), 
\end{eqnarray}
where the potential operator $\hat{V}$ is 
\begin{equation}
\hat{V} = \frac{1}{2} \sum_{{\bf q},{\bf k}',{\bf k}"}  V({\bf q} ) c^\dagger_{{\bf k}' +{\bf q}} c^\dagger_{{\bf k}'' -{\bf q}} c_{{\bf k}'' }  c_{{\bf k}'}.
\end{equation}
Using the fact that  \cite{Tuan_Exc_2024}
\begin{equation}
 \hat{V} \,\, c^\dagger_{\bf k} c^\dagger_{\bf p} |0; 0 \rangle    = \sum_{{\bf q}}V({\bf q})  c^\dagger_{\bf k+q} c^\dagger_{\bf p -q} |0; 0 \rangle,
\end{equation}
we get
\begin{eqnarray}
 \langle 0 ;0 | c_{{\bf p}'} c_{{\bf k}'} \,\, \hat{V} \,\, c^\dagger_{\bf k} c^\dagger_{\bf p} |0; 0 \rangle    &=& \sum_{{\bf q}}V({\bf q})  \,\,\,  \langle 0 ;0 | c_{{\bf p}'} c_{{\bf k}'} \,\  c^\dagger_{\bf k+q} c^\dagger_{\bf p -q} |0; 0 \rangle \nonumber \\
  &=& \sum_{{\bf q}}V({\bf q})  \,\,\, \left(  \delta_{{\bf p}',{\bf p -q}} \delta_{{\bf k}', {\bf k+q} }   - \delta_{{\bf p}', {\bf k+q} } \delta_{{\bf k}', {\bf p -q} } \right).
\end{eqnarray}
The first term in parenthesis contributes to the direct scattering which we have considered in Appendix \ref{app:MaDir}. Here, we calculate the exchange matrix element  coming from the second term which has opposite sign
 \begin{eqnarray}
X_{\{n_1,n_2\}; \{k_y, p_y\} }^{\{n'_1,n'_2\};  \{k'_y, p'_y\} } &=& - (i)^{\Delta n_1 + \Delta n_2} \left( \frac{    2 \pi \ell_\text{B} }{ L_x  } \right)^2 \!\!\!\! \sum_{k_x,k'_x;  p_x,p'_x}    \!\!\!   \!\!\! e^{ \mp i \left(k_x k_y - k'_x k'_y  \right) \ell^2_\text{B}}\,\,\,   \tilde{H}_{n_1}(k_x \ell_\text{B})\,\,\,   \tilde{H}^*_{n'_1}(k'_x \ell_\text{B})   \nonumber \\ 
&\times& \,\,\,\sum_{{\bf q}}V({\bf q})  \,\,\ \delta_{{\bf p}', {\bf k+q} } \delta_{{\bf k}', {\bf p -q} }   \,\,\,\,\,\,\,\,  e^{ \mp i \left(p_x p_y - p'_x p'_y  \right) \ell^2_\text{B}} \,\,\,  \tilde{H}_{n_2}(p_x \ell_\text{B}) \,\,\,  \tilde{H}^*_{n'_2}(p'_x \ell_\text{B})    \nonumber \\
&=& - (i)^{\Delta n_1 + \Delta n_2} \left( \frac{    2 \pi \ell_\text{B} }{ L_x  } \right)^2 \sum_{{\bf q}}V({\bf q})   \sum_{k_x;  p_x}    e^{ \mp i \left(k_x k_y - (p_x -q_x) (p_y -q_y)  \right) \ell^2_\text{B}}\,\,\, e^{ \mp i \left(p_x p_y - (k_x+q_x) (k_y+q_y)  \right) \ell^2_\text{B}}  \nonumber \\ 
&\times&    \tilde{H}_{n_1}(k_x \ell_\text{B})\,\,\,   \tilde{H}^*_{n'_1}\left((p_x -q_x) \ell_\text{B}\right)  \,\,\,  \tilde{H}_{n_2}(p_x \ell_\text{B}) \,\,\,  \tilde{H}^*_{n'_2}\left((k_x+q_x) \ell_\text{B}\right)   \,\,\ \delta_{ p'_y, k_y+q_y } \delta_{ k'_y,  p_y -q_y }.    
\end{eqnarray}
Unlike the case of two distinguishable particles in Appendix \ref{app:MaDir}, we consider here the interaction between two identical particles, meaning that the $\pm$ signs for charge type (electron or hole) in the phase factors are the same.  Joining their phase factors  and rearranging terms with $\tilde{H}_n$ yield 
 \begin{eqnarray}
X_{\{n_1,n_2\}; \{k_y, p_y\} }^{\{n'_1,n'_2\};  \{k'_y, p'_y\} }
&=& - (i)^{\Delta n_1 + \Delta n_2} \left( \frac{    2 \pi \ell_\text{B} }{ L_x  } \right)^2 \sum_{{\bf q}}V({\bf q})\,\,\, e^{ \pm 2i  q_x q_y    \ell^2_\text{B}}   \sum_{k_x;  p_x}    e^{  \pm i \left(  (k_y -p_y)q_x   + (k_x-p_x) q_y  \right) \ell^2_\text{B} }  \nonumber \\ 
&\times&    \tilde{H}_{n_1}(k_x \ell_\text{B})\,\,\,  \tilde{H}^*_{n'_2}\left((k_x+q_x) \ell_\text{B}\right) \,\,\,  \tilde{H}_{n_2}(p_x \ell_\text{B})  \,\,\,    \tilde{H}^*_{n'_1}\left((p_x -q_x) \ell_\text{B}\right)     \,\, \delta_{ p'_y, k_y+q_y } \delta_{ k'_y,  p_y -q_y }    \nonumber \\
&=& - \sum_{\bf q} V({\bf q})  \,\,     e^{ \pm i  k_y q_x    \ell^2_\text{B}}  \,\,\, \,\,  (i)^{n'_2-n_1 }\,\, e^{ \pm i  q_x q_y    \ell^2_\text{B}}  \,\,\, \frac{    2 \pi \ell_\text{B} }{ L_x  } \sum_{k_x}   e^{ \pm i k_xq_y   \ell^2_\text{B}} \,\,\,  \tilde{H}_{n_1}\left(k_x \ell_\text{B}\right)  \,\,\,  \tilde{H}^*_{n'_2}\left((k_x +q_x) \ell_\text{B}\right)  \nonumber \\ 
&\times&    (i)^{n'_1-n_2 } e^{ \mp i   p_yq_x   \ell^2_\text{B}}\,\,  e^{ \pm i  q_x q_y    \ell^2_\text{B}}    \,\,\, \frac{    2 \pi \ell_\text{B} }{ L_x  } \sum_{p_x}  e^{ \mp i p_x q_y \ell^2_\text{B}} \, \tilde{H}_{n_2}\left(p_x \ell_\text{B}\right)  \,  \tilde{H}^*_{n'_1}\left((p_x-q_x) \ell_\text{B}\right)  \, \delta_{ p'_y, k_y+q_y } \delta_{ k'_y,  p_y -q_y }     \nonumber \\
&=& -\sum_{\bf q} e^{\pm i(k_y - p_y)q_x \ell_\text{B}^2} \,\, W_{n_1;n_2 }^{n'_2;n'_1 }({\bf q}) \,\,\,  \delta_{ p'_y, k_y+q_y } \delta_{ k'_y,  p_y -q_y }    \nonumber \\
&=& - V_{n_1, k_y;n_2,p_y}^{n'_2,p'_y; n'_1,k'_y }.
\end{eqnarray}

\section{Hamiltonian matrix for exciton in magnetic field and selection rule} \label{app:Hamil}
 Eq.~(\ref{Eq:Secular1}) is equivalent to
\begin{equation}
\sum_{n_\text{e},n_\text{h}, k_\text{e}}  \phi^{K}_{n_\text{e},n_\text{h}} \left( k_\text{e}\right)\,\,\, \left( \varepsilon_{n_\text{e}} + \varepsilon_{n_\text{h}} + V\left({\bf r}_{\text{e}} -  {\bf r}_{\text{h}} \right) \right) \,\,    \left| n_\text{e},k_\text{e}; n_\text{h},K- k_\text{e}\right\rangle  = E^K  \sum_{n_\text{e},n_\text{h}, k_\text{e}}  \phi^{K}_{n_\text{e},n_\text{h}} \left( k_\text{e}\right)\,\,\,  \left| n_\text{e},k_\text{e}; n_\text{h},K- k_\text{e}\right\rangle. 
\end{equation}
Projecting both sides of the equation on the state $\left| n'_\text{e},k'_\text{e}; n'_\text{h},K- k'_\text{e}\right\rangle $ leads to
\begin{equation}
\sum_{n_\text{e},n_\text{h}, k_\text{e}}   \left( \left[\varepsilon_{n_\text{e}} + \varepsilon_{n_\text{h}} \right]\delta_{n_\text{e},n'_\text{e}} \delta_{n_\text{h},n'_\text{h}} \delta_{k_\text{e},k'_\text{e}} + V_{n_\text{e},k_\text{e};n_\text{h},K-k_\text{e} }^{n'_\text{e},k'_\text{e}; n'_\text{h},K-k'_\text{e} }   \right) \,\,    \phi^{K}_{n_\text{e},n_\text{h}} \left( k_\text{e}\right)   = E^K  \,\,\,  \phi^{K}_{n'_\text{e},n'_\text{h}} \left( k'_\text{e}\right). \label{Eq:BHIni}
\end{equation}
Using the Coulomb matrix element in Eq.~(\ref{Eq:MatrixC}) and summing over $k'_\text{e}$ on both sides of the equation, we arrive at the equation for the contracted exciton envelope function, $\phi^{K}_{n_\text{e},n_\text{h}}   = \sum_{ k_\text{e}}\phi^{K}_{n_\text{e},n_\text{h}} \left( k_\text{e}\right) $,
\begin{eqnarray}
    \left[\varepsilon_{n'_\text{e}} + \varepsilon_{n'_\text{h}} \right]  \phi^{K}_{n'_\text{e},n'_\text{h}}  +  \sum_{n_\text{e},n_\text{h}, k_\text{e},k'_\text{e}} \left(  \sum_{\bf q} e^{iKq_x \ell_\text{B}^2} \,\, W_{n_\text{e};n_\text{h} }^{n'_\text{e};n'_\text{h} }({\bf q}) \delta_{k'_\text{e},k_\text{e}+q_y}    \right) \,\,    \phi^{K}_{n_\text{e},n_\text{h}} \left( k_\text{e}\right)   &=& E^K  \,\,\,  \phi^{K}_{n'_\text{e},n'_\text{h}}  \nonumber \\
    \left[\varepsilon_{n'_\text{e}} + \varepsilon_{n'_\text{h}} \right]  \phi^{K}_{n'_\text{e},n'_\text{h}}  +  \sum_{n_\text{e},n_\text{h}, k_\text{e}} \left(  \sum_{{\bf q},k'_\text{e}} e^{iKq_x \ell_\text{B}^2} \,\, W_{n_\text{e};n_\text{h} }^{n'_\text{e};n'_\text{h} }({\bf q}) \delta_{k'_\text{e},k_\text{e}+q_y}    \right) \,\,    \phi^{K}_{n_\text{e},n_\text{h}} \left( k_\text{e}\right)   &=& E^K  \,\,\,  \phi^{K}_{n'_\text{e},n'_\text{h}}  \nonumber \\
    \left[\varepsilon_{n'_\text{e}} + \varepsilon_{n'_\text{h}} \right]  \phi^{K}_{n'_\text{e},n'_\text{h}}  +  \sum_{n_\text{e},n_\text{h}, k_\text{e}} \left(  \sum_{\bf q} e^{iKq_x \ell_\text{B}^2} \,\, W_{n_\text{e};n_\text{h} }^{n'_\text{e};n'_\text{h} }({\bf q})    \right) \,\,    \phi^{K}_{n_\text{e},n_\text{h}} \left( k_\text{e}\right)   &=& E^K  \,\,\,  \phi^{K}_{n'_\text{e},n'_\text{h}}  \nonumber \\
    \left[\varepsilon_{n'_\text{e}} + \varepsilon_{n'_\text{h}} \right]  \phi^{K}_{n'_\text{e},n'_\text{h}}  +  \sum_{n_\text{e},n_\text{h}} \left(  \sum_{\bf q} e^{iKq_x \ell_\text{B}^2} \,\, W_{n_\text{e};n_\text{h} }^{n'_\text{e};n'_\text{h} }({\bf q})    \right) \,\,    \phi^{K}_{n_\text{e},n_\text{h}}    &=& E^K  \,\,\,  \phi^{K}_{n'_\text{e},n'_\text{h}}  \nonumber \\
 \sum_{n_\text{e},n_\text{h}} H_{n_\text{e},n_\text{h} }^{n'_\text{e},n'_\text{h} }  \,\,\,  \phi^{K}_{n_\text{e},n_\text{h}}   &=& E^K  \,\,\,  \phi^{K}_{n'_\text{e},n'_\text{h}},
\label{app:ExEigenEq}
\end{eqnarray}
 where the Hamiltonian matrix $H_{n_\text{e},n_\text{h} }^{n'_\text{e},n'_\text{h} }$  is given by
 \begin{equation}
 H_{n_\text{e},n_\text{h} }^{n'_\text{e},n'_\text{h} }  =    \left[\varepsilon_{n'_\text{e}} + \varepsilon_{n'_\text{h}} \right] \delta_{n_\text{e},n'_\text{e}} \delta_{n_\text{h},n'_\text{h}}  + \widetilde{V}_{n_\text{e},n_\text{h} }^{n'_\text{e},n'_\text{h} }.
 \end{equation}
The effective potential matrix element is defined as
\begin{equation}
\widetilde{V}_{n_\text{e},n_\text{h} }^{n'_\text{e},n'_\text{h} } =  \sum_{\bf q} e^{iKq_x \ell_\text{B}^2} \,\, W_{n_\text{e};n_\text{h} }^{n'_\text{e};n'_\text{h} }({\bf q}). 
\end{equation}
Using Eqs.$\,$(\ref{Eq:Wq}) and (\ref{Eq:Sq}), we get
\begin{eqnarray}
\widetilde{V}_{n_\text{e},n_\text{h} }^{n'_\text{e},n'_\text{h} } &=&  \sum_{\bf q} e^{iKq_x \ell_\text{B}^2} \,\,  V({\bf q}) \,\, S_{n_\text{e}}^{n'_\text{e}}({\bf q}) \,\, S_{n_\text{h}}^{n'_\text{h}}({- \bf q}) \nonumber \\
&=&i^{|\Delta n_\text{e}| -|\Delta n_\text{h}|  }\sum_{\bf q} e^{iKq_x \ell_\text{B}^2} \,\,  V({\bf q}) \,\,  e^{i(\Delta n_\text{e} - \Delta n_\text{h})\theta} \,\,\,  \tilde{L}^{|\Delta n_\text{e}|}_{\tilde{n}_\text{e}} \left( \frac{q^2 \ell_\text{B}^2}{2} \right)   \,\,\,  \tilde{L}^{|\Delta n_\text{h}|}_{\tilde{n}_\text{h}} \left( \frac{q^2 \ell_\text{B}^2}{2} \right)\nonumber \\
&=&i^{|\Delta n_\text{e}| -|\Delta n_\text{h}| } \frac{A}{4\pi^2} \int_0^{\infty}  \tilde{L}^{|\Delta n_\text{e}|}_{\tilde{n}_\text{e}} \left( \frac{q^2 \ell_\text{B}^2}{2} \right)   \,\,\,  \tilde{L}^{|\Delta n_\text{h}|}_{\tilde{n}_\text{h}} \left( \frac{q^2 \ell_\text{B}^2}{2} \right) q dq    \int_0^{2 \pi}  e^{iKq \ell_\text{B}^2 \cos \theta} \,\,  V({\bf q}) \,\,  e^{i(\Delta n_\text{e} - \Delta n_\text{h})\theta}
d\theta  \,. \,  
\label{app:Wt}
\end{eqnarray}
If the Coulomb potential has circular symmetry, $V({\bf q}) \equiv V(q)$, it can be moved out of the angle integral which  then can be   performed analytically. We consider two such cases for zero- and finite-momentum excitons 
\subsection{Zero-momentum (exciton in the light cone)}
When $K = 0$, the angle integral in Eq.~(\ref{app:Wt}) is 
\begin{equation}
 \int_0^{2 \pi}  e^{i(\Delta n_\text{e} - \Delta n_\text{h})\theta} d\theta \,\,\, = \,\,\, 2 \pi \,\, \delta_{\Delta n_\text{e},  \Delta n_\text{h}} .
 \label{app:del}
\end{equation}
This selection rule means that scattering between the electron and hole LLs satisfies 
\begin{equation}
n'_\text{e} - n_\text{e} = n'_\text{h} - n_\text{h}.
 \label{app:pair}
\end{equation}
The matrix element then becomes
\begin{eqnarray}
\widetilde{V}_{n_\text{e},n_\text{h} }^{n'_\text{e},n'_\text{h} } 
&=& \frac{A}{2\pi} \int_0^{\infty}  \tilde{L}^{|\Delta n|}_{\tilde{n}_\text{e}} \left( \frac{q^2 \ell_\text{B}^2}{2} \right)   \,\,\,  \tilde{L}^{|\Delta n|}_{\tilde{n}_\text{h}} \left( \frac{q^2 \ell_\text{B}^2}{2} \right) q  V(q) \,\,   dq  \nonumber  \\
&=& -  \int_0^{\infty}  \tilde{L}^{|\Delta n|}_{\tilde{n}_\text{e}} \left( \frac{q^2 \ell_\text{B}^2}{2} \right)   \,\,\,  \tilde{L}^{|\Delta n|}_{\tilde{n}_\text{h}} \left( \frac{q^2 \ell_\text{B}^2}{2} \right)  \frac{ e^2}{  \epsilon(q)} \,\,   dq   \nonumber \\
&=& -  \frac{ e^2}{ \ell_\text{B}}  \int_0^{\infty}   \tilde{L}^{|\Delta n|}_{\tilde{n}_\text{e}} \left( \frac{x^2}{2} \right)   \,\,\,  \tilde{L}^{|\Delta n|}_{\tilde{n}_\text{h}} \left( \frac{x^2}{2} \right)  \frac{ dx}{  \epsilon(x/\ell_\text{B})} \,\,,\,      
\label{App:WnnK0}
\end{eqnarray}
where $\Delta n_\text{e} =  \Delta n_\text{h} \equiv \Delta n$, and $\epsilon(q)$ is the dielectric screening function.

\subsection{ Finite-momentum exciton, $K \ne 0$}

 In this case, the angle integral can still be performed analytically by using the following identity for any integer $m$
\begin{equation}
\int_0^{2 \pi} e^{iz \cos \theta} e^{im \theta} d \theta = 2 \pi i^m J_m(z). 
\end{equation} 
 The matrix element then becomes
\begin{eqnarray}
\widetilde{V}_{n_\text{e},n_\text{h} }^{n'_\text{e},n'_\text{h} } 
&=&i^{|\Delta n_\text{e}| + \Delta n_\text{e} -|\Delta n_\text{h}|  -\Delta n_\text{h} } \frac{A}{2\pi} \int_0^{\infty}  \tilde{L}^{|\Delta n_\text{e}|}_{\tilde{n}_\text{e}} \left( \frac{q^2 \ell_\text{B}^2}{2} \right)   \,\,\,  \tilde{L}^{|\Delta n_\text{h}|}_{\tilde{n}_\text{h}} \left( \frac{q^2 \ell_\text{B}^2}{2} \right) q  V(q) \,\,  J_{\Delta^*_n} \left(Kq \ell_\text{B}^2 \right)  dq  \nonumber \\
&=&- i^{|\Delta n_\text{e}| + \Delta n_\text{e} -|\Delta n_\text{h}|  -\Delta n_\text{h} }\,\, \frac{e^2}{\ell_\text{B}} \int_0^{\infty}  \tilde{L}^{|\Delta n_\text{e}|}_{\tilde{n}_\text{e}} \left( \frac{x^2}{2} \right)   \,\,\,  \tilde{L}^{|\Delta n_\text{h}|}_{\tilde{n}_\text{h}} \left( \frac{x^2}{2} \right)  \frac{ J_{\Delta^*_n} \left(K \ell_\text{B} x \right)}{ \epsilon(x/\ell_\text{B})} \,\,    dx  ,
\label{App:WnnKf}
\end{eqnarray}
\end{widetext}
where $\Delta^*_n = \Delta n_\text{e} - \Delta n_\text{h}$. The matrix element is real  because $(|\Delta n_\text{e}| + \Delta n_\text{e} -|\Delta n_\text{h}|  -\Delta n_\text{h} )$ is an even number. For small $K$, the contribution from scattering with  $\Delta^*_n = 0$ is still significant compared to contributions from $\Delta^*_n \neq 0$ since $J_{\Delta^*_n }(0) = \delta_{\Delta^*_n,0}$ (all Bessel functions $J_n(x)$ start from $0$ at $x=0$ except for $J_0(x)$ which starts from 1).

\section{Properties of $\tilde{L}^m_n(x)$} \label{app:LFunc}

Calculations in the main text frequently use the functions $\tilde{L}^m_n(x)$ where $ x\ge 0$.   The functions, which are related to the generalized Laguerre polynomials $L^m_n(x)$ through
\begin{equation}
\tilde{L}^m_n(x) = \sqrt{  \frac{ n! }{    \left( n+ m \right)! } \,\,\,  x^{m} \,\,  e^{-x  }  }       \,\,\,\,     L^{m}_{n}\left( x\right), 
\label{app:DefLnm}
\end{equation}
can be classified into different classes, each with the same $m$. The functions of each class are orthonormal
\begin{equation}
\int_0^{\infty} \tilde{L}^m_n(x) \tilde{L}^m_k (x) dx = \delta_{n,k} .
\end{equation}
The advantage of using $\tilde{L}^m_n(x)$ in numerical calculations over the generalized Laguerre polynomials $L^m_n(x)$  is that $\tilde{L}^m_n(x)$ is  numerically controllable (i.e., it does not blow up fast across the domain $x \ge 0$). Another advantage of  $\tilde{L}^m_n(x)$  is that their numerical values can be obtained from a recurrence relation with the initial conditions 
\begin{eqnarray}
n=0\!&:& \,\, \tilde{L}_0^m(x)  = \sqrt{  \frac{ 1 }{    m ! } \,\,\,  x^{m} \,\,  e^{-x  }  }\,\,,   \\
n=1\!&:& \,\, \tilde{L}_1^m(x)  = \sqrt{  \frac{ 1 }{    \left(  m +1 \right)! } \,\,\,  x^{m} \,\,  e^{-x  }  }       \,\,   \left( 1+m - x\right),\,\,\,\,\,\,\,\, \nonumber
\end{eqnarray}
and linear recurrence for $n\ge 2$,
\begin{eqnarray}  
  \tilde{L}_n^m(x)  \,\,&=& \,\, \frac{ (2n-1+m - x)}{\sqrt{n(n+m)}}     \,\,\,  \tilde{L}_{n-1}^m(x)     \nonumber \\   & - &   \sqrt{  \frac{ (n-1)\left( n+ m -1\right)  }{  n  \left( n+ m \right)  }}  \,\,\,    \tilde{L}_{n-2}^m(x)  .     
\end{eqnarray}
These relations facilitate the calculation of $\tilde{L}_n^m(x)$ for large values of $n$. 

\begin{figure*}[t] 
\centering
\includegraphics[width=16cm]{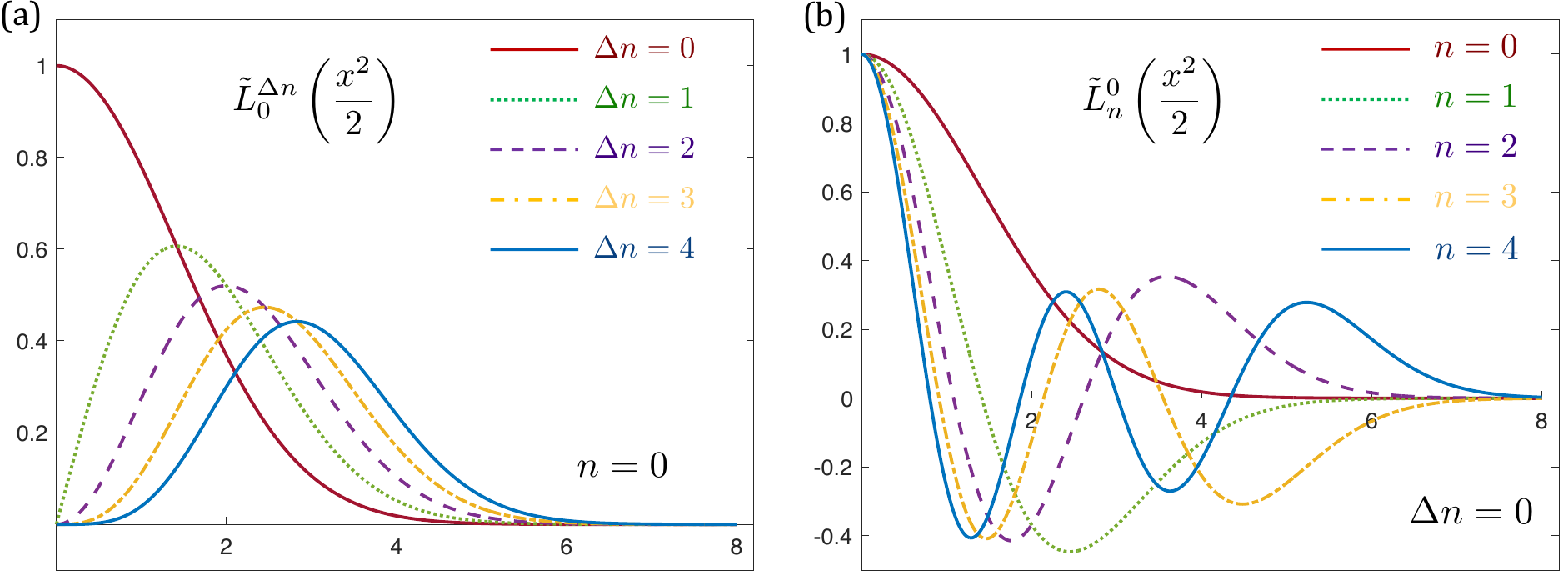}
\caption{ (a) and (b) The five lowest order functions $\tilde{L}_0^{\Delta n}\left(x^2/2\right)$ and $\tilde{L}_n^0\left(x^2/2\right)$, respectively.}\label{fig:Lnm} 
\end{figure*}

The integrands in Eqs.$\,$(\ref{App:WnnK0}) and (\ref{App:WnnKf}) comprise the functions $\tilde{L}_n^{\Delta n}\left(x^2/2\right)$. The case of $\tilde{L}_{0}^{\Delta n}\left(x^2/2\right)$, shown in Fig.~\ref{fig:Lnm}(a), is related to scattering with the lowest LL ($n=0$ or $n' = 0$). Recalling that $x =q \ell_\text{B}$, the integrand reaches a maximum value at
 \begin{equation}
q_0  = \sqrt{2 \Delta n}/\ell_\text{B},
 \end{equation}
where $q_0$ is closely related to the effective transferred momentum for scattering processes that involve the lowest LL. The case of $\tilde{L}^0_n\left(x^2/2\right)$, shown in Fig. \ref{fig:Lnm}(b), contributes to intra-LL scattering ($\Delta n = 0$).  Comparing Figs.~\ref{fig:Lnm}(a) and (b), we notice that
\begin{equation}
\tilde{L}^{\Delta n}_n(x=0) = \delta_{\Delta n,0},
\end{equation}   
which together with the fact that the Coulomb potential is strongest when $q \rightarrow 0$ (or $x \rightarrow 0$), explain why intra-LL scattering processes ($\Delta n =0$) are more effective 
than inter-LL scattering processes ($\Delta n \neq 0$). 

\begin{figure}
\centering
\includegraphics[width=7.5cm]{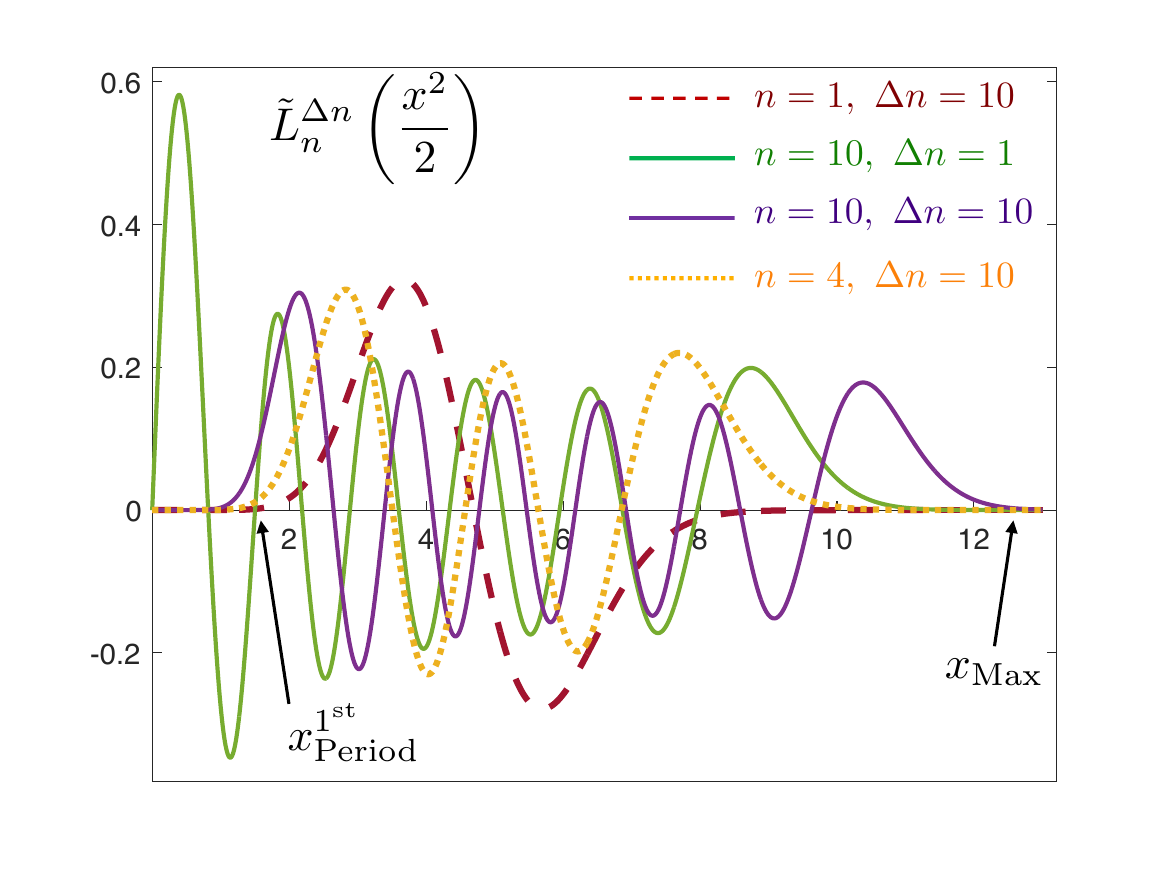}
\caption{Few examples of $\tilde{L}^{\Delta n}_n\left(x^2/2\right)$ for $ n \neq 0$ and $\Delta n \neq 0$.}\label{fig:LnmN0} 
\end{figure}

Figure~\ref{fig:Lnm}(b) shows that $\tilde{L}^0_n\left(x^2/2\right)$ oscillates above and below the $x$-axis, having $n$ zeros before it eventually vanishes at large $x$. This behavior stems from the $n$ solutions of the generalized Laguerre polynomials $L^0_n(x)$ in Eq.~(\ref{app:DefLnm}).  The oscillatory pattern of $\tilde{L}^{\Delta n}_n\left(x^2/2\right)$ persists in cases that $n \ne 0$ and $\Delta n \ne 0$, as shown in Fig.~\ref{fig:LnmN0}, rendering the numerical calculations difficult. In fact, the reason we did not change the integration variable from $x$ to $u=x^2/2$ in Eq.$\,$(\ref{App:WnnK0}) is that the oscillations are relatively regular (i.e., spaced out) when working with $\tilde{L}^{\Delta n}_n\left(x^2/2\right)$, thereby allowing us to use a less dense mesh when carrying the numerical integration. Even with that choice, the integration in Eq.~(\ref{App:WnnK0}) when $n$ is relatively large remains more time consuming than diagonalizing the Hamiltonian.

To perform the numerical integration, we identify the oscillatory period and the function domain. For a certain $n$, the shortest period is typically the first period of $\tilde{L}^{1}_n\left(x^2/2\right)$, indicated by the arrow labeled $x^{1^\text{st}}_\text{Period}$ in Fig.~\ref{fig:LnmN0} for $n = 10$. The integration cutoff has to be large enough to cover the region in which $\tilde{L}^{\Delta n}_n\left(x^2/2\right)$ is not negligible. This cutoff is largest when $\Delta n=n$, indicated by the arrow labeled $x_\text{Max}$ in Fig.~\ref{fig:LnmN0}. To guarantee accurate results, the number of mesh points in the  numerical integration has to be a few orders of magnitude larger than $x_\text{Max}/x^{1^\text{st}}_\text{Period}$. Table A lists the values of $x_\text{Max}$ and $x^{1^\text{st}}_\text{Period}$ for different values of $n$. 

\begin{widetext}
\begin{tabularx}{\textwidth} { 
  | >{\raggedright\arraybackslash}X 
  | >{\centering\arraybackslash}X 
  | >{\centering\arraybackslash}X 
  | >{\centering\arraybackslash}X 
  | >{\centering\arraybackslash}X 
  | >{\centering\arraybackslash}X 
  | >{\centering\arraybackslash}X 
  | >{\centering\arraybackslash}X 
  | >{\centering\arraybackslash}X 
  | >{\centering\arraybackslash}X 
  | >{\raggedleft\arraybackslash}X | }
 \hline
 Table A & $n =300$ & $n =200$ & $n =100$ & $n =90$ & $n =70$ &  $n= 50$ &  $n= 40$ &  $n= 30$ &  $n= 20$ &  $n = 10$ \\
 \hline
 $x_\text{Max}$ & 62 & 52 & 37 & 35 & 31  & 27 & 25 &  22 &19 & 15  \\
\hline
 $x^{1^\text{st}}_\text{Period}$ & 0.285 & 0.35& 0.494 & 0.522 & 0.592  & 0.699 & 0.780 &  0.894 &1.10 & 1.55  \\
\hline
\end{tabularx}
\end{widetext}
 

\section{Exciton excited states in 2D materials} \label{app:HighStates}
To further validate the model, we compare the calculated exciton energies with those calculated through the real-space Hamiltonian of an exciton with angular momentum $L_z = \hbar l$. We first consider the case of zero magnetic field, where the Schr\"{o}dinger equation $\hat{H} \phi({\bf r}) = E\,\, \phi({\bf r})$ is derived from the Hamiltonian (${\bf r}$ is the relative motion coordinate of the exciton),
\begin{equation}
\hat{H}= 
 -\frac{\hbar^2 }{2 \mu } \nabla^2 + V({\bf r}), 
 \label{Eq:HalOrb}
\end{equation}
where $\mu^{-1}= m_\text{e}^{-1} + m_\text{h}^{-1}$. Writing the kinetic energy operator in polar coordinates, $ -\hbar^2 \nabla^2 /(2 \mu) = \hat{K}_r + \hat{K}_\theta$, the radial component reads
\begin{equation}
\hat{K}_r =  -\frac{\hbar^2 }{2 \mu }  \frac{1}{r} \frac{\partial}{\partial r} \left( r \frac{\partial}{\partial r} \right), 
\label{Eq:RaKr}
\end{equation}
and the angular component reads
\begin{equation}
\hat{K}_\theta =  -\frac{\hbar^2 }{2 \mu }  \frac{1}{r^2} \frac{\partial^2}{\partial \theta^2}  = \frac{\hat{L}_z^2}{2 \mu r^2}.  \label{Eq:RaKtheta}
\end{equation}
Since the Coulomb potential has radial symmetry, $ V({\bf r}) \equiv  V(r)$,  the Hamiltonian commutes with the exciton angular momentum operator $\hat{L}_z = -i \hbar \partial /\partial \theta$, i.e., $\left[ \hat{H}, \hat{L}_z\right] = 0 $. Consequently, the radial and angular components of the wavefunction can be separated,  $\phi({\bf r}) = \phi_l(r) e^{i l \theta}/\sqrt{2 \pi}$, where the radial component satisfies
\begin{eqnarray}
\left[ \frac{\hbar^2 }{2 \mu } \left( \! \frac{l^2}{r^2} - \frac{\partial^2}{\partial r^2}  -  \frac{1}{r}  \frac{\partial}{\partial r} \! \right)  \!+\! V(r)\right]  \phi_l(r) = E_l \phi_l(r). \label{Eq:SchroEqmne0} \,\,\,\,\,\,\,\,\,\,\, \,\,\,\,
\end{eqnarray} 
Solving this equation for $l= \{0, 1, 2, 3, ...\}$, we find the eigenstates with $\{s, p, d, f, ...\}$ character. 

Using the stochastic variational method (SVM) \cite{Varga_CPC_2008,VanTuan_PRB22, VanTuan_SVM22}, we express the radial eigenfunction that corresponds to angular momentum  $\hbar l $ as a superposition of correlated Gaussian basis functions
\begin{equation}
\phi_l(r) = \sum_{i} c^l_i \varphi^l_i(r), \label{eq:svm_corr}
\end{equation}
where
\begin{equation}
\varphi^l_i(r) = r^l e^{-\frac{1}{2} \alpha_i r^2}.
\label{app:Gauss}
\end{equation} 
The asymptotic behavior $\varphi^l_i(r) \sim  r^l$ when $r \rightarrow 0$ stems from the centrifugal potential, $\hbar^2l^2/2\mu r^2$ in Eq.~(\ref{Eq:SchroEqmne0}), which dominates near the origin. $\alpha_i$ is a variational parameter chosen to minimize the energy $E_l$ through the trisection method  \cite{VanTuan_SVM22}.  The  $c^l_i$ parameters in Eq.~(\ref{eq:svm_corr}) are obtained from the secular equation
\begin{equation}
\sum_{i} H_{ij} c^l_{j} = E_l  \sum_{i} O_{ij}  c^l_j. 
\end{equation}
The overlap matrix is defined as 
\begin{equation}
O_{ij} =  \int_0^{\infty}     \varphi^l_i(r)  \varphi^l_j(r)  \,\, r dr =  \frac{l!}{ 2 \alpha_{ij}^{l+1}}\,,
\end{equation}
where $\alpha_{ij} = (\alpha_i + \alpha_j)/2$.

The Hamiltonian matrix $H_{ij}  = K_{ij}  + F_{ij} + V_{ij}  $ includes the kinetic energy matrix element
\begin{eqnarray}
K_{ij} &=& -\frac{\hbar^2 }{2 \mu } \int_0^{\infty}     \varphi^l_i(r)  \left( \frac{\partial^2}{\partial r^2}  +  \frac{1}{r}  \frac{\partial}{\partial r}  \right)  \varphi^l_j(r)  \,\, r dr \nonumber \\
&=&  \frac{\hbar^2 }{2 \mu }\,\, \frac{  (l+1) \alpha_i \alpha_j - l\alpha_{ij}^2   }{  \alpha_{ij}} \,\, O_{ij},
\end{eqnarray}
a matrix element associated with the centrifugal force
\begin{eqnarray}
F_{ij} &=& \frac{\hbar^2 }{2 \mu } \int_0^{\infty}    \varphi^l_i(r)   \,\, \frac{l^2}{r^2}   \,\,   \varphi^l_j(r)  \,\, r dr =  \frac{\hbar^2 l }{2 \mu } \alpha_{ij} \,\, O_{ij},\,\,\,\,\,\,\,\,\,\,\,\,\,\,\,\,\,
\end{eqnarray}
and the potential energy matrix element
\begin{equation}
V_{ij} =  \int_0^{\infty}   \varphi^l_i(r) V(r) \varphi^l_j(r) \,\, r dr.
\end{equation}
Due to translation symmetry, it is convenient to write the potential in momentum space  
\begin{equation}
V({\bf r}) = \sum_{\bf q} V(q) e^{i{\bf q.r}} ,
\end{equation}
where the Fourier transform of the potential is given by
\begin{equation}
V(q) = \frac{2 \pi e^2}{A\,\, q \,\, \epsilon(q)}.
\end{equation}
The potential matrix element involves dielectric screening contribution from the environment surrounding the 2D semiconductor through $\epsilon(q)$, yielding 
\begin{equation} 
V_{ij} =  
 - 2e^2  \sqrt{\alpha_{ij}}  \, O_{ij}   \int_0^{\infty}       \frac{ e^{-x^2} }{  \epsilon( 2\sqrt{\alpha_{ij}} x) } \,\, L_l(x^2)  \, dx\,,
\end{equation}
where $ L_l(x) $ is the Laguerre polynomial. Using the Rytova-Keldysh model to study transition-metal dichalcogenide monolayers \cite{Rytova_MSU67,Keldysh_JETP79},
$\epsilon(q) = \epsilon_v (1 + r_0 q)$, where $\epsilon_v$ is  the averaged dielectric constant of the media below and above the monolayer, and $r_0= 2 \pi \alpha /\epsilon_v$ is related to the polarizability $\alpha$ of the monolayer \cite{Cudazzo_PRB2011}, we get 
\begin{eqnarray} 
V_{ij} =   - \frac{e^2}{ \epsilon_v r_0 } \,\, O_{ij}   \,\,   \int_0^{\infty}       \frac{ e^{-x^2} }{  \beta_{ij}+  x} \,\, L_l(x^2)  \, dx \,,
\end{eqnarray}  
where $\beta_{ij} = \left( 2 r_0\sqrt{\alpha_{ij}} \right)^{-1} $.

\begin{table}[t!]
\caption{\label{tab:table2} Energies and state labels of the five lowest-energy exciton states of hBN-encapsulated WSe$_2$, calculated with the Rytova-Keldysh dielectric screening at zero magnetic field.  Parameters for the calculations can be found in Appendix \ref{App:Para}.}
\begin{ruledtabular}
\begin{tabular}{c|cccccc}
  &Ground& 1$^{\text{st}}$  & 2$^\text{nd}$ &3$^\text{rd}$ &4$^\text{th}$ &5$^\text{th}$ \\ \hline 
 $E_l$  (meV) &$- 171.5$ & $ - 53.1$  &$- 40.9 $ & $-22.9 $ & $-21.2$ & $-17.8$  \\ 
 State labels  &1$s$     & 2$p$        &2$s$       & 3$d$       & 3$p$ & 3$s$    \\ 
\end{tabular}
\end{ruledtabular}
\end{table}

\begin{table}[t!]
\caption{\label{tab:table3} Same as Table \ref{tab:table2} but for $\epsilon_v = 4.4, r_0 $ \cite{Donck_PRB18}.}
\begin{ruledtabular}
\begin{tabular}{c|cccccc}
  &Ground& 1$^{\text{st}}$  & 2$^\text{nd}$ &3$^\text{rd}$ &4$^\text{th}$ &5$^\text{th}$ \\ \hline 
 This work &$- 148.0$ & $ - 42.1$  &$- 32.7 $ & $-17.4 $ & $-16.4$ & $-13.9$  \\ 
 Ref. \cite{Donck_PRB18} &$- 146$ & $ - 45$  &$- 34 $ & $-19 $ & $-18$ & $-15$  \\ 
 State labels  &1$s$     & 2$p$        &2$s$       & 3$d$       & 3$p$ & 3$s$    \\ 
\end{tabular}
\end{ruledtabular}
\end{table}

\begin{table}[t!]
\caption{\label{tab:table4} The comparison between exciton energies for $s$ states obtained by this work and the ones in  Ref.~\cite{Spiridonova_PLA20}. The parameters used in both cases are $\epsilon_v = 3.97, r_0=1.26$ nm, and $\mu=0.2$ \cite{Spiridonova_PLA20,Liu_PRB19}.}
\begin{ruledtabular}
\begin{tabular}{c|cccc}
 State labels &$1s$& $2s$  & $3s$ &$4s$  \\ \hline 
 This work &$- 172.401$ & $ - 43.839$  &$- 19.538 $ & $-10.966 $  \\ 
 Ref. \cite{Spiridonova_PLA20} &$- 172.38$ & $ - 43.84$  &$- 19.54 $ & $-10.97 $   \\ 
\end{tabular}
\end{ruledtabular}
\end{table}

Table \ref{tab:table2} shows the obtained results from these SVM calculations, using $m_\text{e} = 0.29m_0$, $m_\text{h} = 0.36m_0$ and $\epsilon_v= 3.8$ \cite{VanTuan_PRB22}. These results reinforce the finding in Table~\ref{tab:table1}. Furthermore, the SVM results match those calculated in Ref.~\cite{Donck_PRB18} with $\epsilon_v= 4.4$, as shown in Table \ref{tab:table3}. The differences between the two sets of results are less than 3~meV, although we use quadratic energy dispersion Hamiltonion in the effective mass model while the authors of Ref. \cite{Donck_PRB18} have used a Dirac Hamiltonian.   Particularly, Ref.~\cite{Spiridonova_PLA20} used the same quadratic energy dispersion Hamiltonian but with different parameters for   $\mu$, $\epsilon_v$,   and $r_0$. Table~\ref{tab:table4} shows the calculated energies of the $s$-state  when using our SVM model with the same parameters used in Ref.~\cite{Spiridonova_PLA20}. The perfect agreement reinforces the derived method in this work.    


\section{Exciton Hamiltonian in magnetic field}\label{App:ExHalMag}
The centrifugal potential in Eq.~(\ref{Eq:SchroEqmne0}), $\hbar^2 l^2 / 2 \mu r^2$, is the same for $\pm l$, meaning that their energy levels are degenerate at zero magnetic field ($E_{l} = E_{-l}$). The degeneracy comes from the equivalence between excitons with angular momentum pointing upward and downward with respect to the 2D plane. This degeneracy is lifted under the effect of an out-of-plane magnetic field due to its interaction with the exciton magnetic moment generated by an exciton with finite angular momentum ($l \neq 0$). To study this interaction, we first write the real-space Hamiltonian of the exciton, 
\begin{eqnarray}
\hat{\mathcal{H}} &=& \frac{1}{2 m_\text{e} } \left( - i \hbar \nabla_\text{e}  + e {\bf A}_{{\bf r}_\text{e}} \right)^2 + \frac{1}{2 m_\text{h} } \left( - i \hbar \nabla_\text{h}  - e {\bf A}_{{\bf r}_\text{h}} \right)^2 \nonumber \\
&+& V({\bf r}_\text{e} - {\bf r}_\text{h} ).
\end{eqnarray}
Switching to the center-of-mass and relative motion coordinates, ${\bf R} = \left( m_\text{e} {\bf r}_\text{e} +  m_\text{h} {\bf r}_\text{h} \right)/(m_\text{e} + m_\text{h})$ and ${\bf r} = {\bf r}_\text{e} -  {\bf r}_\text{h}$, the Hamiltonian can be rewritten as
\begin{widetext}
\begin{eqnarray}
\hat{\mathcal{H}}=  \frac{ P^2}{2M}     + \frac{e}{M} {\bf A}_{\bf r}. {\bf P} + \frac{ p^2}{2\mu}+\frac{e}{\mu} {\bf A}_{\bf R}. {\bf p} +  \left( \frac{e}{m_\text{e}}-\frac{e}{m_\text{h}}\right) {\bf A}_{\bf r}. \left( {\bf p}  + e  {\bf A}_{\bf R}\right) +  \frac{e^2 {\bf A}^2_{\bf R}}{2 \mu} +  \frac{e^2}{2} \left( \frac{M}{m_\text{e} m_\text{h}}-\frac{3}{M}\right) {\bf A}^2_{\bf r} + V(r)\,,   \,\,\,\,\, \,\,\,\,\,\,\,\, \,\,\,
\end{eqnarray}
\end{widetext}
where ${\bf P} = -i \hbar \nabla_{\bf R}$ is the center-of-mass momentum operator, ${\bf p} = -i \hbar \nabla_{\bf r}  $ is the relative motion momentum operator, $M = m_\text{e} + m_\text{h}$ is the translational mass of the exciton, and we have used
\begin{eqnarray}
\nabla_{\text{e}} = \nabla_{\bf r} + \frac{m_\text{e}}{M}  \nabla_{\bf R} \quad,\quad  \nabla_{\text{h}} = -\nabla_{\bf r} + \frac{m_\text{h}}{M}  \nabla_{\bf R}\,\, .\qquad
\end{eqnarray}

The Hamiltonian can be simplified by using the canonical transformation \cite{Pico_PRB25,Stier_PRL2018,Lamb1952,KnoxBook,MiuraBook,Cong_EMO18,Donck_PRB18,Raczynska_NJP19,Gorkov_JETP68,Whittaker_SSC88, Katsch_PRB20,Chang_JCP22,Aleksandrov_PRB24,Pico_PRB25,Spiridonova_PLA20} 
\begin{equation}
\Psi'= \Psi \exp \left( - \frac{i }{\hbar}e    {\bf A_R} . {\bf r}  \right),
\end{equation}
and  $  {\bf A}_{{\bf r}} . {\bf R}   = -{\bf A}_{{\bf R}} . {\bf r} $. The new Hamiltonian is 
\begin{eqnarray}
\hat{\mathcal{H}'} &=&\frac{{\bf P}^2}{2M} +  \frac{{\bf p}^2}{2\mu} + V(r) \nonumber \\ &+& 2\frac{e}{M}  {\bf A}_{\bf r}  . {\bf P}   +  \frac{2\mu^{\text{ex}}_\text{B}}{\hbar} {\bf A}_{\bf r}. {\bf p}  +   \frac{e^2{\bf A}^2_{\bf r}}{2 \mu}\,\,,\qquad
\end{eqnarray}
where $\mu^{\text{ex}}_\text{B} = e \hbar/2 m_\text{e} - e \hbar/2 m_\text{h}$ is the exciton Bohr magneton.  The total hamiltonian $\hat{\mathcal{H}'}$ commutes with $\bf P$, $\left[ \hat{\mathcal{H}'}, {\bf P}\right] = 0 $, indicating that the eigenvector of $\bf P$, $e^{i{\bf K.R}}/\sqrt{A}$, diagonalizes $\hat{\mathcal{H}'}$ as well.   The total wave function is therefore
\begin{equation}
\Psi({\bf R,r}) = \frac{e^{i{\bf K.R}}}{\sqrt{A}} \psi({\bf r}) ,
\end{equation}
where the Hamiltonian for $\psi({\bf r})$ is
\begin{eqnarray}
\hat{\mathcal{H}'}_{\bf K}
  &=&\frac{{\bf K}^2}{2M} +  \frac{{\bf p}^2}{2\mu}  + V(r)    \nonumber \\ &+& 2\frac{e\hbar }{M}  {\bf A}_{\bf r}  . {\bf K}   + \frac{2\mu^{\text{ex}}_\text{B}}{\hbar} {\bf A}_{\bf r}. {\bf p}  +   \frac{e^2{\bf A}^2_{\bf r}}{2 \mu} \,\,. \qquad
\end{eqnarray}
Considering an exciton in the light-cone ($K=0$), the dependence on translational motion drops, and the Hamiltonian only depends on the relative motion
\begin{eqnarray}
\hat{H}  =  -\frac{\hbar^2 }{2 \mu } \nabla^2 + V(r)       +\frac{2\mu^{\text{ex}}_\text{B}}{\hbar} {\bf A}_{\bf r}. {\bf p}  +   \frac{e^2{\bf A}^2_{\bf r}}{2 \mu}.  \,\,  \,\,\,\,\,\, \,\,\,\,
\end{eqnarray}
The Hamiltonian is similar to Eq.$\,$(\ref{Eq:HalOrb}) with two additional terms due to the vector potential. Using the symmetric gauge, ${\bf A}_{\bf r} = \left(  {\bf B} \times {\bf r} \right)/2$, and the definition of the angular momentum  operator $\hat{{\bf L}}=({\bf r}\times  {\bf p} )$, the first term can be rewritten as
 \begin{eqnarray}
    \frac{2\mu^{\text{ex}}_\text{B}}{\hbar} {\bf A}_{\bf r}. {\bf p} = \frac{\mu^{\text{ex}}_\text{B}}{\hbar} {\bf B} .({\bf r}\times  {\bf p} ) =  \frac{\mu^{\text{ex}}_\text{B}}{\hbar} B \hat{L}_z.
\end{eqnarray}
This term denotes the interaction of an out-of-plane magnetic field,  ${\bf B} = B{\hat{\textbf{z}}}$, with the exciton angular momentum.  The second term  leads to a diamagnetic shift due to the  confinement potential caused by the magnetic field \cite{Stier_PRL2018}
\begin{equation}
 \frac{e^2{\bf A}^2_{\bf r}}{2 \mu} = \frac{e^2 B^2 r^2}{8 \mu}.
\end{equation} 
The Hamiltonian becomes
\begin{eqnarray}
\hat{H}
  =  -\frac{\hbar^2 }{2 \mu } \nabla^2 + V(r)       +   \frac{\mu^{\text{ex}}_\text{B}}{\hbar} B\hat{L}_z    +  \frac{e^2 B^2 r^2}{8 \mu}.   
\end{eqnarray}
Using Eqs.~(\ref{Eq:RaKr})-(\ref{Eq:RaKtheta}) to express $ -\hbar^2 \nabla^2 /(2 \mu) $ in polar coordinates, we arrive at Eq.~(\ref{r-Hamil}) of the main text 
\begin{equation}
\hat{H}
  =  \hat{K}_r  +  \frac{\hat{L}_z^2}{2 \mu r^2} +  V(r)  +   \mu^{\text{ex}}_\text{B}  B  \frac{\hat{L}_z}{\hbar}  +    \frac{e^2 B^2 r^2}{8 \mu}   .
\end{equation}


\section{Extension of the model for quasi-2D and 3D systems}\label{App:3DExt}
In this section we study the cases of  quasi-2D systems with confinement potential $V_\perp(z)$ and 3D systems without confinement potential ($V_\perp(z) = 0$). The Hamiltonian in Eq.(\ref{eq:H}) becomes
\begin{equation}
H_0 = \frac{1}{2m} \left( p_x^2 + \left(p_y \mp eBx \right)^2 + p_z^2 \right) + V^{\mp}_\perp(z). \label{eq:H3D}
\end{equation}
The upper (lower) sign denotes the electron (hole) case. Assuming that $V^{\mp}_\perp(z)$is independent of $x$ and $y$, the  eigenvector in Eq.~(\ref{Eq:Eivec}) can be written as
\begin{equation}
\psi_{k_y}(x,y) = \frac{e^{ik_yy}}{\sqrt{L_y}}f_{k_y}(x) \,\,\, \chi^{\mp}_{n_z}(z), \label{Eq:Eivec1}
\end{equation}
where $n_z$ is the discrete index of the trapping levels and   $\chi^{\mp}_{n_z}(z)$ is the eigenvector of the $z$-direction Hamiltonian $H_z = p_z^2 /2m   + V^{\mp}_\perp(z)$  with the corresponding eigenenergy $\xi^{\mp}_{n_z}$
\begin{equation}
\left(\frac{p_z^2 }{2m}   + V^{\mp}_\perp(z)\right) \chi^{\mp}_{n_z}(z)  = \xi^{\mp}_{n_z} \,\,  \chi^{\mp}_{n_z}(z). \label{eq:H3Dz}
\end{equation}
The total eigen-energies in Eq.~(\ref{Eq:EigEner}) become 
\begin{equation} 
\varepsilon_{n,k_y,n_z} = \hbar \omega \left( n+ \frac{1}{2} \right) +  \xi^{\mp}_{n_z}.  
\label{Eq:EigEner1}
\end{equation}
In the  3D case  ($V^{\mp}_\perp(z) = 0$),  the discrete index due to the confinement  along the $z$ direction, $n_z$, becomes $k_z$. The corresponding energy and eigenvector in this case become $\xi^{\mp}_{k_z} = k_z^2/2m$ and $\chi^{\mp}_{k_z}(z) = e^{ik_zz}/\sqrt{L_z}$, respectively, where $L_z$ is sample size along the $z$ direction.

The derivation for the potential matrix element, which now becomes $V_{n_c,k,n_{z_c};n_d,p,n_{z_d} }^{n'_c,k',n'_{z_c};n'_d,p',n'_{z_d} }$, can be performed by following Eq.~{\ref{Eq.VDeri}}, in which $ \langle {\bf k}';{\bf p}' | \hat{V} | {\bf k};{\bf p} \rangle  $ is replaced by 
\begin{widetext}
\begin{eqnarray}
\langle {\bf k}',n'_{z_c};{\bf p}',n'_{z_d}  | \hat{V} | {\bf k},n_{z_c};{\bf p},n_{z_d} \rangle  &=&  \sum_{\bf q} V({\bf q})\,\,\,  Z_{n_{z_c}}^{n'_{z_c}}(q_z) \,\, Z_{n_{z_d}}^{n'_{z_d}}(-q_z)  \,\,\,  \delta_{ k'_x, k_x+q_x} \delta_{ k'_y, k_y + q_y}  \delta_{ p'_x, p_x-q_x} \delta_{ p'_y, p_y - q_y}    
\end{eqnarray}
where 
\begin{equation}
Z_{n_z}^{n'_z}(q_z) =   \int    \chi^{\mp^*}_{n'_z} (z)  e^{i q_z z}  \chi^{\mp}_{n_z}(z)  \,\, dz 
\end{equation}
and ${\bf q} = (q_x,q_y,q_z)$ is the 3D  vector of the transferred momentum. In the 3D case, $Z_{n_z}^{n'_z}(q_z) $ becomes  $Z_{k_z}^{k'_z}(q_z) = \delta_{ k'_z, k_z+q_z} $ which dictates the momentum conservation along the $z$ direction.
The factor $ Z_{n_z}^{n'_z}(q_z) $ is  incorporated into $S_{n}^{n'}({\bf q})$ to make
\begin{equation}
S_{n,n_z}^{n',n'_z}({\bf q})  = S_{n}^{n'}({\bf q}_\parallel) Z_{n_z}^{n'_z}(q_z),
\label{Eq:Sq1}
\end{equation}
 where $S_{n}^{n'}({\bf q}_\parallel)$ is given by Eq.~(\ref{Eq:Sq}) with ${\bf q}_\parallel =(q_x,q_y)$. This leads to a modified form of $W_{n_c;n_d }^{n'_c;n'_d }({\bf q}) $ in Eq.~(\ref{Eq:Wq}) as follows
 \begin{equation}
W_{n_c,n_{z_c};n_d,n_{z_d} }^{n'_c,n'_{z_c};n'_d,n'_{z_d} }({\bf q})  = V({\bf q}) \,\, S_{n_c,n_{z_c}}^{n'_c,n'_{z_c}}({\bf q}) \,\, S_{n_d,n_{z_d}}^{n'_d,n'_{z_d}}({- \bf q}).
\end{equation}
The obtained  potential matrix element (quasi-2D version of Eq.~(\ref{Eq:MatrixC})) is  
\begin{eqnarray}V_{n_c,k,n_{z_c};n_d,p,n_{z_d} }^{n'_c,k',n'_{z_c};n'_d,p',n'_{z_d} }= \!\sum_{\bf q} e^{i(\pm k \mp p)q_x \ell_\text{B}^2} \,\, W_{n_c,n_{z_c};n_d,n_{z_d} }^{n'_c,n'_{z_c};n'_d,n'_{z_d} }({\bf q})   \delta_{k;p}^{k';p'}\!(q_y).  \qquad
\label{Eq:MatrixC1}
\end{eqnarray}

Next, we consider the magneto-exciton problem in quasi-2D and 3D systems. The bound-state wavefunction, given by Eq.~(\ref{eq:phi_b}), now becomes
\begin{equation}
|\Psi_{K,\text{b}} \rangle = \sum_{n_\text{e},n_{z_\text{e}},n_\text{h},n_{z_\text{h}}, k_\text{e}}  \phi^{K}_{n_\text{e},n_{z_\text{e}},n_\text{h},n_{z_\text{h}}}\!\!\left( k_\text{e}\right)\,\,\,  \left| n_\text{e},k_\text{e},n_{z_\text{e}}; n_\text{h},K- k_\text{e},n_{z_\text{h}}\right\rangle, \label{eq:phi_b1}
\end{equation}
where the exciton wavefunction, $\phi^{K}_{n_\text{e},n_{z_\text{e}},n_\text{h},n_{z_\text{h}}}\!\!\left( k_\text{e}\right)$, satisfies the quasi-2D version  of Eqs.~(\ref{Eq:BHIni}) and  (\ref{app:ExEigenEq}),
\begin{eqnarray}
&&\!\!\!\!\!\!\!\!\!\!\!\!\!\!\! \sum_{n_\text{e},n_{z_\text{e}},n_\text{h},n_{z_\text{h}}, k_\text{e}}   \left( \left[\varepsilon_{n_\text{e}} + \varepsilon_{n_\text{h}} +  \xi^{-}_{n_{z_\text{e}}} + \xi^{+}_{n_{z_\text{h}}}  \right]\delta_{n_\text{e},n'_\text{e}} \delta_{n_\text{h},n'_\text{h}}          \delta_{n_{z_\text{e}},n'_{z_\text{e}}} \delta_{n_{z_\text{h}},n'_{z_\text{h}}}            \delta_{k_\text{e},k'_\text{e}} \,\,\,\,\,\, +\,\,\, \,\,\, V_{n_\text{e},k_\text{e},n_{z_\text{e}};n_\text{h},K-k_\text{e},n_{z_\text{h}} }^{n'_\text{e},k'_\text{e},n'_{z_\text{e}}; n'_\text{h},K-k'_\text{e},n'_{z_\text{h}} }   \right) \,\,     \phi^{K}_{n_\text{e},n_{z_\text{e}},n_\text{h},n_{z_\text{h}}}\!\!\left( k_\text{e}\right)    \nonumber \\
&=& E^K  \,\,\,     \phi^{K}_{n'_\text{e},n'_{z_\text{e}},n'_\text{h},n'_{z_\text{h}}}\!\!\left( k'_\text{e}\right)    \nonumber \\
   &&\!\!\! \left[\varepsilon_{n'_\text{e}} + \varepsilon_{n'_\text{h}} + \xi^{-}_{n'_{z_\text{e}}} + \xi^{+}_{n'_{z_\text{h}}}  \right]  \phi^{K}_{n'_\text{e},n'_{z_\text{e}},n'_\text{h},n'_{z_\text{h}}}  +  \!\!\!\sum_{n_\text{e},n_{z_\text{e}},n_\text{h},n_{z_\text{h}}, k_\text{e},k'_\text{e}} \left(  \sum_{\bf q} e^{iKq_x \ell_\text{B}^2} \,\, W_{n_\text{e},n_{z_\text{e}};n_\text{h},n_{z_\text{h}} }^{n'_\text{e},n'_{z_\text{e}};n'_\text{h},n'_{z_\text{h}} }({\bf q}) \delta_{k'_\text{e},k_\text{e}+q_y}    \right) \,\,    \phi^{K}_{n_\text{e},n_{z_\text{e}},n_\text{h},n_{z_\text{h}}} \left( k_\text{e}\right)  \nonumber \\ 
    &=& E^K  \,\,\,  \phi^{K}_{n'_\text{e},n'_{z_\text{e}},n'_\text{h},n'_{z_\text{h}}}  \nonumber \\
       &&\!\!\! \left[\varepsilon_{n'_\text{e}} + \varepsilon_{n'_\text{h}} + \xi^{-}_{n'_{z_\text{e}}} + \xi^{+}_{n'_{z_\text{h}}}  \right]  \phi^{K}_{n'_\text{e},n'_{z_\text{e}},n'_\text{h},n'_{z_\text{h}}}  +  \!\!\!\sum_{n_\text{e},n_{z_\text{e}},n_\text{h},n_{z_\text{h}}, k_\text{e}} \left(  \sum_{{\bf q},k'_\text{e}} e^{iKq_x \ell_\text{B}^2} \,\, W_{n_\text{e},n_{z_\text{e}};n_\text{h},n_{z_\text{h}} }^{n'_\text{e},n'_{z_\text{e}};n'_\text{h},n'_{z_\text{h}} }({\bf q}) \delta_{k'_\text{e},k_\text{e}+q_y}    \right) \,\,    \phi^{K}_{n_\text{e},n_{z_\text{e}},n_\text{h},n_{z_\text{h}}} \left( k_\text{e}\right)  \nonumber \\ 
    &=& E^K  \,\,\,  \phi^{K}_{n'_\text{e},n'_{z_\text{e}},n'_\text{h},n'_{z_\text{h}}}  \nonumber \\
      &&\!\!\! \left[\varepsilon_{n'_\text{e}} + \varepsilon_{n'_\text{h}} + \xi^{-}_{n'_{z_\text{e}}} + \xi^{+}_{n'_{z_\text{h}}}  \right]  \phi^{K}_{n'_\text{e},n'_{z_\text{e}},n'_\text{h},n'_{z_\text{h}}}  +  \!\!\!\sum_{n_\text{e},n_{z_\text{e}},n_\text{h},n_{z_\text{h}}, k_\text{e}} \left(  \sum_{{\bf q}} e^{iKq_x \ell_\text{B}^2} \,\, W_{n_\text{e},n_{z_\text{e}};n_\text{h},n_{z_\text{h}} }^{n'_\text{e},n'_{z_\text{e}};n'_\text{h},n'_{z_\text{h}} }({\bf q})    \right) \,\,    \phi^{K}_{n_\text{e},n_{z_\text{e}},n_\text{h},n_{z_\text{h}}} \left( k_\text{e}\right)  \nonumber \\ 
    &=& E^K  \,\,\,  \phi^{K}_{n'_\text{e},n'_{z_\text{e}},n'_\text{h},n'_{z_\text{h}}}  \nonumber \\   
      &&\!\!\! \left[\varepsilon_{n'_\text{e}} + \varepsilon_{n'_\text{h}} + \xi^{-}_{n'_{z_\text{e}}} + \xi^{+}_{n'_{z_\text{h}}}  \right]  \phi^{K}_{n'_\text{e},n'_{z_\text{e}},n'_\text{h},n'_{z_\text{h}}}  +  \!\!\!\sum_{n_\text{e},n_{z_\text{e}},n_\text{h},n_{z_\text{h}}} \left(  \sum_{{\bf q}} e^{iKq_x \ell_\text{B}^2} \,\, W_{n_\text{e},n_{z_\text{e}};n_\text{h},n_{z_\text{h}} }^{n'_\text{e},n'_{z_\text{e}};n'_\text{h},n'_{z_\text{h}} }({\bf q})    \right) \,\,    \phi^{K}_{n_\text{e},n_{z_\text{e}},n_\text{h},n_{z_\text{h}}}   \nonumber \\ 
    &=& E^K  \,\,\,  \phi^{K}_{n'_\text{e},n'_{z_\text{e}},n'_\text{h},n'_{z_\text{h}}}  \nonumber \\   
&&\!\!\! \!\!\!  \!\!\! \!\!\! \!\!\!  \sum_{n_\text{e},n_{z_\text{e}},n_\text{h},n_{z_\text{h}}}  H_{n_\text{e},n_{z_\text{e}};n_\text{h},n_{z_\text{h}} }^{n'_\text{e},n'_{z_\text{e}};n'_\text{h},n'_{z_\text{h}} }\,\,\,\,\,\,\,  \phi^{K}_{n_\text{e},n_{z_\text{e}},n_\text{h},n_{z_\text{h}}}  \,\,\,\,\,\,= \,\,\,\,\,\,   E^K  \,\,\,  \phi^{K}_{n'_\text{e},n'_{z_\text{e}},n'_\text{h},n'_{z_\text{h}}}.  
\end{eqnarray}
 The Hamiltonian matrix $H_{n_\text{e},n_{z_\text{e}};n_\text{h},n_{z_\text{h}} }^{n'_\text{e},n'_{z_\text{e}};n'_\text{h},n'_{z_\text{h}} }       $  is given by
 \begin{equation}
 H_{n_\text{e},n_{z_\text{e}};n_\text{h},n_{z_\text{h}} }^{n'_\text{e},n'_{z_\text{e}};n'_\text{h},n'_{z_\text{h}} }  =    \left[\varepsilon_{n'_\text{e}} + \varepsilon_{n'_\text{h}} + \xi^{-}_{n'_{z_\text{e}}} + \xi^{+}_{n'_{z_\text{h}}}   \right] \delta_{n_\text{e},n'_\text{e}} \delta_{n_\text{h},n'_\text{h}}   \delta_{n_{z_\text{e}},n'_{z_\text{e}}} \delta_{n_{z_\text{h}},n'_{z_\text{h}}}      + \widetilde{V}_{n_\text{e},n_{z_\text{e}};n_\text{h},n_{z_\text{h}} }^{n'_\text{e},n'_{z_\text{e}};n'_\text{h},n'_{z_\text{h}} }  ,
 \end{equation}
where the effective potential matrix element is defined as
\begin{eqnarray}
\widetilde{V}_{n_\text{e},n_{z_\text{e}};n_\text{h},n_{z_\text{h}} }^{n'_\text{e},n'_{z_\text{e}};n'_\text{h},n'_{z_\text{h}} }   &=&  \sum_{\bf q} e^{iKq_x \ell_\text{B}^2} \,\,  W_{n_\text{e},n_{z_\text{e}};n_\text{h},n_{z_\text{h}} }^{n'_\text{e},n'_{z_\text{e}};n'_\text{h},n'_{z_\text{h}} }({\bf q}) \nonumber \\
&=& \sum_{\bf q} e^{iKq_x \ell_\text{B}^2} \,\,  V({\bf q}) \,\, S_{n_\text{e}}^{n'_\text{e}}({\bf q}_\parallel) \,\, S_{n_\text{h}}^{n'_\text{h}}({- \bf q}_\parallel) \,\,\,\,\, Z_{n_{z_\text{e}}}^{n'_{z_\text{e}}}(q_z) \,\, Z_{n_{z_\text{h}}}^{n'_{z_\text{h}}}(-q_z) \nonumber \\ 
&=&  \sum_{ q_z} Z_{n_{z_\text{e}}}^{n'_{z_\text{e}}}(q_z) \,\, Z_{n_{z_\text{h}}}^{n'_{z_\text{h}}}(-q_z)    \sum_{{\bf q}_\parallel }e^{iKq_x \ell_\text{B}^2} \,\,  V({\bf q}) \,\, S_{n_\text{e}}^{n'_\text{e}}({\bf q}_\parallel) \,\, S_{n_\text{h}}^{n'_\text{h}}({- \bf q}_\parallel).  \label{Eq:Vquasi}
\end{eqnarray}
In the 3D case, the matrix elements become 
\begin{eqnarray}
\widetilde{V}_{n_\text{e},k_{z_\text{e}};n_\text{h},k_{z_\text{h}} }^{n'_\text{e},k'_{z_\text{e}};n'_\text{h},k'_{z_\text{h}} } 
&=&  \sum_{ q_z} \delta_{ k'_{z_\text{e}}, k_{z_\text{e}}+q_z}   \delta_{ k'_{z_\text{h}}, k_{z_\text{h}}- q_z}     \sum_{{\bf q}_\parallel }e^{iKq_x \ell_\text{B}^2} \,\,  V({\bf q}) \,\, S_{n_\text{e}}^{n'_\text{e}}({\bf q}_\parallel) \,\, S_{n_\text{h}}^{n'_\text{h}}({- \bf q}_\parallel).  \label{Eq:V3D}
\end{eqnarray}
\end{widetext}
The two delta functions select $q_z = k'_{z_\text{e}} - k_{z_\text{e}}$ and guarantee  the $z$-direction momentum conservation for the interaction between the electron and hole components. That is,  only  matrix elements with $k_{z_\text{e}}+ k_{z_\text{h}} = k'_{z_\text{e}}  + k'_{z_\text{h}} $ have finite values. 
The sum over ${\bf q}_\parallel$ in Eqs.~(\ref{Eq:Vquasi}) and ~(\ref{Eq:V3D}) can be performed in a similar way to the one in Eq.~(\ref{app:Wt}) from which we also obtain Eq.~(\ref{app:del}) and Eq.~(\ref{app:pair}). 
All in all, the pairing law in Eq.~(\ref{Eq:LLCoupling}) still holds for quasi-2D and 3D systems because the motion along the $z$-direction is independent of the angular momentum parameters (i.e., of the magnetic quantum number).

\section{Parameters}\label{App:Para}

The following parameters are used in  the calculations of magneto-excitons in hBN-encapsulated WSe$_2$ monolayer (Figures~\ref{fig:DEpm}-\ref{fig:Diamag}). 
\begin{enumerate}
\item Effective masses: $m_\text{h} = 0.36 m_0$ and $m_\text{e} = 0.29m_0$  \cite{Kormanyos_2DMater15}. These masses correspond, respectively, to a hole in the topmost valley of the valence band and an electron in the top spin-split valley of the conduction band (optically active valley). 
\item The parameters of the Rytova-Keldysh dielectric function used in Eq.~(\ref{Eq:Vpot}), $\epsilon(u) = \epsilon_v(1 + r_0 u/\ell_B)$, are $r_0 = $ 1.18 nm and  $\epsilon_v =3.8$ \cite{Cai_SSC07,VanTuan_PRB18,Stier_PRL2018,Berkelbach_PRB13}. 
\item The simulations include $N = 200$~LLs. The calculation of exciton states with magnetic quantum number $l$ involves diagonalizing a matrix of size $(N - |l|) \times (N -|l|)$. As an example, it takes less than a minute to calculate the $s$ states ($l=0$) at a given magnetic field on a laptop computer, where most of this time is dedicated to computation of the potential matrix elements.

\end{enumerate}


\end{document}